\documentclass[table]{article}
\usepackage[a4paper, margin=1in]{geometry}
\usepackage{authblk}
\usepackage[utf8]{inputenc}

\usepackage[hyphens]{url}
\usepackage{booktabs}
\usepackage{datatool,graphicx,xcolor,cals}
\usepackage{todonotes}
\usepackage{pgfplots}
\usepackage{pgfplotstable}
\usepackage{tikz}
\usepackage{caption}
\usepackage{subcaption}
\usepackage{listings}
\usepackage{balance}

\usepackage{tcolorbox}
\usepackage{listings}
\lstdefinelanguage{PowerShell}{
	morekeywords={
		Add-Content,Add-PSSnapin,Clear-Content,Clear-History,Clear-Host,Clear-Item,Clear-ItemProperty,Clear-Variable,Compare-Object,Connect-PSSession,ConvertFrom-String,Convert-Path,Copy-Item,Copy-ItemProperty,Disable-PSBreakpoint,Disconnect-PSSession,Enable-PSBreakpoint,Enter-PSSession,Exit-PSSession,Export-Alias,Export-Csv,Export-PSSession,ForEach-Object,Format-Custom,Format-Hex,Format-List,Format-Table,Format-Wide,Get-Alias,Get-ChildItem,Get-Clipboard,Get-Command,Get-ComputerInfo,Get-Content,Get-History,Get-Item,Get-ItemProperty,Get-ItemPropertyValue,Get-Job,Get-Location,Get-Member,Get-Module,Get-Process,Get-PSBreakpoint,Get-PSCallStack,Get-PSDrive,Get-PSSession,Get-PSSnapin,Get-Service,Get-TimeZone,Get-Unique,Get-Variable,Get-WmiObject,Group-Object,help,Import-Alias,Import-Csv,Import-Module,Import-PSSession,Invoke-Command,Invoke-Expression,Invoke-History,Invoke-Item,Invoke-RestMethod,Invoke-WebRequest,Invoke-WmiMethod,Measure-Object,mkdir,Move-Item,Move-ItemProperty,New-Alias,New-Item,New-Module,New-PSDrive,New-PSSession,New-PSSessionConfigurationFile,New-Variable,Out-GridView,Out-Host,Out-Printer,Pop-Location,powershell_ise.exe,Push-Location,Receive-Job,Receive-PSSession,Remove-Item,Remove-ItemProperty,Remove-Job,Remove-Module,Remove-PSBreakpoint,Remove-PSDrive,Remove-PSSession,Remove-PSSnapin,Remove-Variable,Remove-WmiObject,Rename-Item,Rename-ItemProperty,Resolve-Path,Resume-Job,Select-Object,Select-String,Set-Alias,Set-Clipboard,Set-Content,Set-Item,Set-ItemProperty,Set-Location,Set-PSBreakpoint,Set-TimeZone,Set-Variable,Set-WmiInstance,Show-Command,Sort-Object,Start-Job,Start-Process,Start-Service,Start-Sleep,Stop-Job,Stop-Process,Stop-Service,Suspend-Job,Tee-Object,Trace-Command,Wait-Job,Where-Object,Write-Output
	},
	morekeywords={
		Add-AppxPackage,Add-AppxProvisionedPackage,Add-AppxVolume,Add-BitsFile,Add-CertificateEnrollmentPolicyServer,Add-Computer,Add-Content,Add-History,Add-JobTrigger,Add-KdsRootKey,Add-LocalGroupMember,Add-Member,Add-PSSnapin,Add-Type,Add-WindowsCapability,Add-WindowsDriver,Add-WindowsImage,Add-WindowsPackage,Checkpoint-Computer,Clear-Content,Clear-EventLog,Clear-History,Clear-Item,Clear-ItemProperty,Clear-KdsCache,Clear-RecycleBin,Clear-Tpm,Clear-Variable,Clear-WindowsCorruptMountPoint,Compare-Object,Complete-BitsTransfer,Complete-DtiagnosticTransaction,Complete-Transaction,Confirm-SecureBootUEFI,Connect-PSSession,Connect-WSMan,ConvertFrom-Csv,ConvertFrom-Json,ConvertFrom-SecureString,ConvertFrom-String,ConvertFrom-StringData,Convert-Path,Convert-String,ConvertTo-Csv,ConvertTo-Html,ConvertTo-Json,ConvertTo-ProcessMitigationPolicy,ConvertTo-SecureString,ConvertTo-TpmOwnerAuth,ConvertTo-Xml,Copy-Item,Copy-ItemProperty,Debug-Job,Debug-Process,Debug-Runspace,Disable-AppBackgroundTaskDiagnosticLog,Disable-ComputerRestore,Disable-JobTrigger,Disable-LocalUser,Disable-PSBreakpoint,Disable-PSRemoting,Disable-PSSessionConfiguration,Disable-RunspaceDebug,Disable-ScheduledJob,Disable-TlsCipherSuite,Disable-TlsEccCurve,Disable-TlsSessionTicketKey,Disable-TpmAutoProvisioning,Disable-WindowsErrorReporting,Disable-WindowsOptionalFeature,Disable-WSManCredSSP,Disconnect-PSSession,Disconnect-WSMan,Dismount-AppxVolume,Dismount-WindowsImage,Enable-AppBackgroundTaskDiagnosticLog,Enable-ComputerRestore,Enable-JobTrigger,Enable-LocalUser,Enable-PSBreakpoint,Enable-PSRemoting,Enable-PSSessionConfiguration,Enable-RunspaceDebug,Enable-ScheduledJob,Enable-TlsCipherSuite,Enable-TlsEccCurve,Enable-TlsSessionTicketKey,Enable-TpmAutoProvisioning,Enable-WindowsErrorReporting,Enable-WindowsOptionalFeature,Enable-WSManCredSSP,Enter-PSHostProcess,Enter-PSSession,Exit-PSHostProcess,Exit-PSSession,Expand-WindowsCustomDataImage,Expand-WindowsImage,Export-Alias,Export-BinaryMiLog,Export-Certificate,Export-Clixml,Export-Console,Export-Counter,Export-Csv,Export-FormatData,Export-ModuleMember,Export-PfxCertificate,Export-ProvisioningPackage,Export-PSSession,Export-StartLayout,Export-StartLayoutEdgeAssets,Export-TlsSessionTicketKey,Export-Trace,Export-WindowsCapabilitySource,Export-WindowsDriver,Export-WindowsImage,Find-Package,Find-PackageProvider,ForEach-Object,Format-Custom,Format-List,Format-SecureBootUEFI,Format-Table,Format-Wide,Get-Acl,Get-Alias,Get-AppxDefaultVolume,Get-AppxPackage,Get-AppxPackageManifest,Get-AppxProvisionedPackage,Get-AppxVolume,Get-AuthenticodeSignature,Get-BitsTransfer,Get-Certificate,Get-CertificateAutoEnrollmentPolicy,Get-CertificateEnrollmentPolicyServer,Get-CertificateNotificationTask,Get-ChildItem,Get-CimAssociatedInstance,Get-CimClass,Get-CimInstance,Get-CimSession,Get-Clipboard,Get-CmsMessage,Get-Command,Get-ComputerInfo,Get-ComputerRestorePoint,Get-Content,Get-ControlPanelItem,Get-Counter,Get-Credential,Get-Culture,Get-DAPolicyChange,Get-Date,Get-DeliveryOptimizationLog,Get-DeliveryOptimizationPerfSnap,Get-DeliveryOptimizationPerfSnapThisMonth,Get-DeliveryOptimizationStatus,Get-DODownloadMode,Get-DOPercentageMaxBackgroundBandwidth,Get-DOPercentageMaxForegroundBandwidth,Get-Event,Get-EventLog,Get-EventSubscriber,Get-ExecutionPolicy,Get-FormatData,Get-Help,Get-History,Get-Host,Get-HotFix,Get-Item,Get-ItemProperty,Get-ItemPropertyValue,Get-Job,Get-JobTrigger,Get-KdsConfiguration,Get-KdsRootKey,Get-LocalGroup,Get-LocalGroupMember,Get-LocalUser,Get-Location,Get-Member,Get-Module,Get-Package,Get-PackageProvider,Get-PackageSource,Get-PfxCertificate,Get-PfxData,Get-PmemDisk,Get-PmemPhysicalDevice,Get-PmemUnusedRegion,Get-Process,Get-ProcessMitigation,Get-ProvisioningPackage,Get-PSBreakpoint,Get-PSCallStack,Get-PSDrive,Get-PSHostProcessInfo,Get-PSProvider,Get-PSReadlineKeyHandler,Get-PSReadlineOption,Get-PSSession,Get-PSSessionCapability,Get-PSSessionConfiguration,Get-PSSnapin,Get-Random,Get-Runspace,Get-RunspaceDebug,Get-ScheduledJob,Get-ScheduledJobOption,Get-SecureBootPolicy,Get-SecureBootUEFI,Get-Service,Get-TimeZone,Get-TlsCipherSuite,Get-TlsEccCurve,Get-Tpm,Get-TpmEndorsementKeyInfo,Get-TpmSupportedFeature,Get-TraceSource,Get-Transaction,Get-TroubleshootingPack,Get-TrustedProvisioningCertificate,Get-TypeData,Get-UICulture,Get-Unique,Get-Variable,Get-WIMBootEntry,Get-WinAcceptLanguageFromLanguageListOptOut,Get-WinCultureFromLanguageListOptOut,Get-WinDefaultInputMethodOverride,Get-WindowsCapability,Get-WindowsDeveloperLicense,Get-WindowsDriver,Get-WindowsEdition,Get-WindowsErrorReporting,Get-WindowsImage,Get-WindowsImageContent,Get-WindowsOptionalFeature,Get-WindowsPackage,Get-WindowsSearchSetting,Get-WinEvent,Get-WinHomeLocation,Get-WinLanguageBarOption,Get-WinSystemLocale,Get-WinUILanguageOverride,Get-WinUserLanguageList,Get-WmiObject,Get-WSManCredSSP,Get-WSManInstance,Group-Object,Import-Alias,Import-BinaryMiLog,Import-Certificate,Import-Clixml,Import-Counter,Import-Csv,Import-LocalizedData,Import-Module,Import-PackageProvider,Import-PfxCertificate,Import-PSSession,Import-StartLayout,Import-TpmOwnerAuth,Initialize-PmemPhysicalDevice,Initialize-Tpm,Install-Package,Install-PackageProvider,Install-ProvisioningPackage,Install-TrustedProvisioningCertificate,Invoke-CimMethod,Invoke-Command,Invoke-CommandInDesktopPackage,Invoke-DscResource,Invoke-Expression,Invoke-History,Invoke-Item,Invoke-RestMethod,Invoke-TroubleshootingPack,Invoke-WebRequest,Invoke-WmiMethod,Invoke-WSManAction,Join-DtiagnosticResourceManager,Join-Path,Limit-EventLog,Measure-Command,Measure-Object,Mount-AppxVolume,Mount-WindowsImage,Move-AppxPackage,Move-Item,Move-ItemProperty,New-Alias,New-CertificateNotificationTask,New-CimInstance,New-CimSession,New-CimSessionOption,New-DtiagnosticTransaction,New-Event,New-EventLog,New-FileCatalog,New-Item,New-ItemProperty,New-JobTrigger,New-LocalGroup,New-LocalUser,New-Module,New-ModuleManifest,New-NetIPsecAuthProposal,New-NetIPsecMainModeCryptoProposal,New-NetIPsecQuickModeCryptoProposal,New-Object,New-PmemDisk,New-ProvisioningRepro,New-PSDrive,New-PSRoleCapabilityFile,New-PSSession,New-PSSessionConfigurationFile,New-PSSessionOption,New-PSTransportOption,New-PSWorkflowExecutionOption,New-ScheduledJobOption,New-SelfSignedCertificate,New-Service,New-TimeSpan,New-TlsSessionTicketKey,New-Variable,New-WebServiceProxy,New-WindowsCustomImage,New-WindowsImage,New-WinEvent,New-WinUserLanguageList,New-WSManInstance,New-WSManSessionOption,Optimize-AppxProvisionedPackages,Optimize-WindowsImage,Out-Default,Out-File,Out-GridView,Out-Host,Out-Null,Out-Printer,Out-String,Pop-Location,Protect-CmsMessage,Publish-DscConfiguration,Push-Location,Read-Host,Receive-DtiagnosticTransaction,Receive-Job,Receive-PSSession,Register-ArgumentCompleter,Register-CimIndicationEvent,Register-EngineEvent,Register-ObjectEvent,Register-PackageSource,Register-PSSessionConfiguration,Register-ScheduledJob,Register-WmiEvent,Remove-AppxPackage,Remove-AppxProvisionedPackage,Remove-AppxVolume,Remove-BitsTransfer,Remove-CertificateEnrollmentPolicyServer,Remove-CertificateNotificationTask,Remove-CimInstance,Remove-CimSession,Remove-Computer,Remove-Event,Remove-EventLog,Remove-Item,Remove-ItemProperty,Remove-Job,Remove-JobTrigger,Remove-LocalGroup,Remove-LocalGroupMember,Remove-LocalUser,Remove-Module,Remove-PmemDisk,Remove-PSBreakpoint,Remove-PSDrive,Remove-PSReadlineKeyHandler,Remove-PSSession,Remove-PSSnapin,Remove-TypeData,Remove-Variable,Remove-WindowsCapability,Remove-WindowsDriver,Remove-WindowsImage,Remove-WindowsPackage,Remove-WmiObject,Remove-WSManInstance,Rename-Computer,Rename-Item,Rename-ItemProperty,Rename-LocalGroup,Rename-LocalUser,Repair-WindowsImage,Reset-ComputerMachinePassword,Resolve-DnsName,Resolve-Path,Restart-Computer,Restart-Service,Restore-Computer,Resume-BitsTransfer,Resume-Job,Resume-ProvisioningSession,Resume-Service,Save-Help,Save-Package,Save-WindowsImage,Select-Object,Select-String,Select-Xml,Send-DtiagnosticTransaction,Send-MailMessage,Set-Acl,Set-Alias,Set-AppBackgroundTaskResourcePolicy,Set-AppxDefaultVolume,Set-AppXProvisionedDataFile,Set-AuthenticodeSignature,Set-BitsTransfer,Set-CertificateAutoEnrollmentPolicy,Set-CimInstance,Set-Clipboard,Set-Content,Set-Culture,Set-Date,Set-DODownloadMode,Set-DOPercentageMaxBackgroundBandwidth,Set-DOPercentageMaxForegroundBandwidth,Set-DscLocalConfigurationManager,Set-ExecutionPolicy,Set-Item,Set-ItemProperty,Set-JobTrigger,Set-KdsConfiguration,Set-LocalGroup,Set-LocalUser,Set-Location,Set-PackageSource,Set-ProcessMitigation,Set-PSBreakpoint,Set-PSDebug,Set-PSReadlineKeyHandler,Set-PSReadlineOption,Set-PSSessionConfiguration,Set-ScheduledJob,Set-ScheduledJobOption,Set-SecureBootUEFI,Set-Service,Set-StrictMode,Set-TimeZone,Set-TpmOwnerAuth,Set-TraceSource,Set-Variable,Set-WinAcceptLanguageFromLanguageListOptOut,Set-WinCultureFromLanguageListOptOut,Set-WinDefaultInputMethodOverride,Set-WindowsEdition,Set-WindowsProductKey,Set-WindowsSearchSetting,Set-WinHomeLocation,Set-WinLanguageBarOption,Set-WinSystemLocale,Set-WinUILanguageOverride,Set-WinUserLanguageList,Set-WmiInstance,Set-WSManInstance,Set-WSManQuickConfig,Show-Command,Show-ControlPanelItem,Show-EventLog,Show-WindowsDeveloperLicenseRegistration,Sort-Object,Split-Path,Split-WindowsImage,Start-BitsTransfer,Start-DscConfiguration,Start-DtiagnosticResourceManager,Start-Job,Start-Process,Start-Service,Start-Sleep,Start-Transaction,Start-Transcript,Stop-Computer,Stop-DtiagnosticResourceManager,Stop-Job,Stop-Process,Stop-Service,Stop-Transcript,Suspend-BitsTransfer,Suspend-Job,Suspend-Service,Switch-Certificate,Tee-Object,Test-Certificate,Test-ComputerSecureChannel,Test-Connection,Test-DscConfiguration,Test-FileCatalog,Test-KdsRootKey,Test-ModuleManifest,Test-Path,Test-PSSessionConfigurationFile,Test-WSMan,Trace-Command,Unblock-File,Unblock-Tpm,Undo-DtiagnosticTransaction,Undo-Transaction,Uninstall-Package,Uninstall-ProvisioningPackage,Uninstall-TrustedProvisioningCertificate,Unprotect-CmsMessage,Unregister-Event,Unregister-PackageSource,Unregister-PSSessionConfiguration,Unregister-ScheduledJob,Unregister-WindowsDeveloperLicense,Update-FormatData,Update-Help,Update-List,Update-TypeData,Update-WIMBootEntry,Use-Transaction,Use-WindowsUnattend,Wait-Debugger,Wait-Event,Wait-Job,Wait-Process,Where-Object,Write-Debug,Write-Error,Write-EventLog,Write-Host,Write-Information,Write-Output,Write-Progress,Write-Verbose,Write-Warning
	},
	morekeywords={
		Add-BitLockerKeyProtector,Add-DnsClientNrptRule,Add-DtcClusterTMMapping,Add-EtwTraceProvider,Add-InitiatorIdToMaskingSet,Add-MpPreference,Add-NetEventNetworkAdapter,Add-NetEventPacketCaptureProvider,Add-NetEventProvider,Add-NetEventVFPProvider,Add-NetEventVmNetworkAdapter,Add-NetEventVmSwitch,Add-NetEventVmSwitchProvider,Add-NetEventWFPCaptureProvider,Add-NetIPHttpsCertBinding,Add-NetLbfoTeamMember,Add-NetLbfoTeamNic,Add-NetNatExternalAddress,Add-NetNatStaticMapping,Add-NetSwitchTeamMember,Add-Odbsn,Add-PartitionAccessPath,Add-PhysicalDisk,Add-Printer,Add-PrinterDriver,Add-PrinterPort,Add-StorageFaultDomain,Add-TargetPortToMaskingSet,Add-VirtualDiskToMaskingSet,Add-VpnConnection,Add-VpnConnectionRoute,Add-VpnConnectionTriggerApplication,Add-VpnConnectionTriggerDnsConfiguration,Add-VpnConnectionTriggerTrustedNetwork,AfterAll,AfterEach,Assert-MockCalled,Assert-VerifiableMocks,Backup-BitLockerKeyProtector,BackupToAAD-BitLockerKeyProtector,BeforeAll,BeforeEach,Block-FileShareAccess,Block-SmbShareAccess,Clear-BitLockerAutoUnlock,Clear-Disk,Clear-DnsClientCache,Clear-FileStorageTier,Clear-Host,Clear-PcsvDeviceLog,Clear-StorageDiagnosticInfo,Close-SmbOpenFile,Close-SmbSession,Compress-Archive,Configuration,Connect-IscsiTarget,Connect-VirtualDisk,Context,convert,ConvertFrom-SddlString,Copy-NetFirewallRule,Copy-NetIPsecMainModeCryptoSet,Copy-NetIPsecMainModeRule,Copy-NetIPsecPhase1AuthSet,Copy-NetIPsecPhase2AuthSet,Copy-NetIPsecQuickModeCryptoSet,Copy-NetIPsecRule,Debug-FileShare,Debug-MMAppPrelaunch,Debug-StorageSubSystem,Debug-Volume,Describe,Disable-BitLocker,Disable-BitLockerAutoUnlock,Disable-DAManualEntryPointSelection,Disable-Dsebug,Disable-MMAgent,Disable-NetAdapter,Disable-NetAdapterBinding,Disable-NetAdapterChecksumOffload,Disable-NetAdapterEncapsulatedPacketTaskOffload,Disable-NetAdapterIPsecOffload,Disable-NetAdapterLso,Disable-NetAdapterPacketDirect,Disable-NetAdapterPowerManagement,Disable-NetAdapterQos,Disable-NetAdapterRdma,Disable-NetAdapterRsc,Disable-NetAdapterRss,Disable-NetAdapterSriov,Disable-NetAdapterVmq,Disable-NetDnsTransitionConfiguration,Disable-NetFirewallRule,Disable-NetIPHttpsProfile,Disable-NetIPsecMainModeRule,Disable-NetIPsecRule,Disable-NetNatTransitionConfiguration,Disable-NetworkSwitchEthernetPort,Disable-NetworkSwitchFeature,Disable-NetworkSwitchVlan,Disable-OdbcPerfCounter,Disable-PhysicalDiskIdentification,Disable-PnpDevice,Disable-PSTrace,Disable-PSWSManCombinedTrace,Disable-ScheduledTask,Disable-SmbDelegation,Disable-StorageEnclosureIdentification,Disable-StorageEnclosurePower,Disable-StorageHighAvailability,Disable-StorageMaintenanceMode,Disable-WdacBidTrace,Disable-WSManTrace,Disconnect-IscsiTarget,Disconnect-VirtualDisk,Dismount-DiskImage,Enable-BitLocker,Enable-BitLockerAutoUnlock,Enable-DAManualEntryPointSelection,Enable-Dsebug,Enable-MMAgent,Enable-NetAdapter,Enable-NetAdapterBinding,Enable-NetAdapterChecksumOffload,Enable-NetAdapterEncapsulatedPacketTaskOffload,Enable-NetAdapterIPsecOffload,Enable-NetAdapterLso,Enable-NetAdapterPacketDirect,Enable-NetAdapterPowerManagement,Enable-NetAdapterQos,Enable-NetAdapterRdma,Enable-NetAdapterRsc,Enable-NetAdapterRss,Enable-NetAdapterSriov,Enable-NetAdapterVmq,Enable-NetDnsTransitionConfiguration,Enable-NetFirewallRule,Enable-NetIPHttpsProfile,Enable-NetIPsecMainModeRule,Enable-NetIPsecRule,Enable-NetNatTransitionConfiguration,Enable-NetworkSwitchEthernetPort,Enable-NetworkSwitchFeature,Enable-NetworkSwitchVlan,Enable-OdbcPerfCounter,Enable-PhysicalDiskIdentification,Enable-PnpDevice,Enable-PSTrace,Enable-PSWSManCombinedTrace,Enable-ScheduledTask,Enable-SmbDelegation,Enable-StorageEnclosureIdentification,Enable-StorageEnclosurePower,Enable-StorageHighAvailability,Enable-StorageMaintenanceMode,Enable-WdacBidTrace,Enable-WSManTrace,Expand-Archive,Export-ODataEndpointProxy,Export-ScheduledTask,Find-Command,Find-DscResource,Find-Module,Find-NetIPsecRule,Find-NetRoute,Find-RoleCapability,Find-Script,Flush-EtwTraceSession,Format-Hex,Format-Volume,Get-AppBackgroundTask,Get-AppxLastError,Get-AppxLog,Get-AutologgerConfig,Get-BitLockerVolume,Get-ClusteredScheduledTask,Get-DAClientExperienceConfiguration,Get-DAConnectionStatus,Get-DAEntryPointTableItem,Get-DedupProperties,Get-Disk,Get-DiskImage,Get-DiskStorageNodeView,Get-DnsClient,Get-DnsClientCache,Get-DnsClientGlobalSetting,Get-DnsClientNrptGlobal,Get-DnsClientNrptPolicy,Get-DnsClientNrptRule,Get-DnsClientServerAddress,Get-DscConfiguration,Get-DscConfigurationStatus,Get-DscLocalConfigurationManager,Get-DscResource,Get-Dtc,Get-DtcAdvancedHostSetting,Get-DtcAdvancedSetting,Get-DtcClusterDefault,Get-DtcClusterTMMapping,Get-Dtefault,Get-DtcLog,Get-DtcNetworkSetting,Get-DtcTransaction,Get-DtcTransactionsStatistics,Get-DtcTransactionsTraceSession,Get-DtcTransactionsTraceSetting,Get-EtwTraceProvider,Get-EtwTraceSession,Get-FileHash,Get-FileIntegrity,Get-FileShare,Get-FileShareAccessControlEntry,Get-FileStorageTier,Get-InitiatorId,Get-InitiatorPort,Get-InstalledModule,Get-InstalledScript,Get-IscsiConnection,Get-IscsiSession,Get-IscsiTarget,Get-IscsiTargetPortal,Get-IseSnippet,Get-LogProperties,Get-MaskingSet,Get-MMAgent,Get-MockDynamicParameters,Get-MpComputerStatus,Get-MpPreference,Get-MpThreat,Get-MpThreatCatalog,Get-MpThreatDetection,Get-NCSIPolicyConfiguration,Get-Net6to4Configuration,Get-NetAdapter,Get-NetAdapterAdvancedProperty,Get-NetAdapterBinding,Get-NetAdapterChecksumOffload,Get-NetAdapterEncapsulatedPacketTaskOffload,Get-NetAdapterHardwareInfo,Get-NetAdapterIPsecOffload,Get-NetAdapterLso,Get-NetAdapterPacketDirect,Get-NetAdapterPowerManagement,Get-NetAdapterQos,Get-NetAdapterRdma,Get-NetAdapterRsc,Get-NetAdapterRss,Get-NetAdapterSriov,Get-NetAdapterSriovVf,Get-NetAdapterStatistics,Get-NetAdapterVmq,Get-NetAdapterVMQQueue,Get-NetAdapterVPort,Get-NetCompartment,Get-NetConnectionProfile,Get-NetDnsTransitionConfiguration,Get-NetDnsTransitionMonitoring,Get-NetEventNetworkAdapter,Get-NetEventPacketCaptureProvider,Get-NetEventProvider,Get-NetEventSession,Get-NetEventVFPProvider,Get-NetEventVmNetworkAdapter,Get-NetEventVmSwitch,Get-NetEventVmSwitchProvider,Get-NetEventWFPCaptureProvider,Get-NetFirewallAddressFilter,Get-NetFirewallApplicationFilter,Get-NetFirewallInterfaceFilter,Get-NetFirewallInterfaceTypeFilter,Get-NetFirewallPortFilter,Get-NetFirewallProfile,Get-NetFirewallRule,Get-NetFirewallSecurityFilter,Get-NetFirewallServiceFilter,Get-NetFirewallSetting,Get-NetIPAddress,Get-NetIPConfiguration,Get-NetIPHttpsConfiguration,Get-NetIPHttpsState,Get-NetIPInterface,Get-NetIPseospSetting,Get-NetIPsecMainModeCryptoSet,Get-NetIPsecMainModeRule,Get-NetIPsecMainModeSA,Get-NetIPsecPhase1AuthSet,Get-NetIPsecPhase2AuthSet,Get-NetIPsecQuickModeCryptoSet,Get-NetIPsecQuickModeSA,Get-NetIPsecRule,Get-NetIPv4Protocol,Get-NetIPv6Protocol,Get-NetIsatapConfiguration,Get-NetLbfoTeam,Get-NetLbfoTeamMember,Get-NetLbfoTeamNic,Get-NetNat,Get-NetNatExternalAddress,Get-NetNatGlobal,Get-NetNatSession,Get-NetNatStaticMapping,Get-NetNatTransitionConfiguration,Get-NetNatTransitionMonitoring,Get-NetNeighbor,Get-NetOffloadGlobalSetting,Get-NetPrefixPolicy,Get-NetQosPolicy,Get-NetRoute,Get-NetSwitchTeam,Get-NetSwitchTeamMember,Get-NetTCPConnection,Get-NetTCPSetting,Get-NetTeredoConfiguration,Get-NetTeredoState,Get-NetTransportFilter,Get-NetUDPEndpoint,Get-NetUDPSetting,Get-NetworkSwitchEthernetPort,Get-NetworkSwitchFeature,Get-NetworkSwitchGlobalData,Get-NetworkSwitchVlan,Get-Odbriver,Get-Odbsn,Get-OdbcPerfCounter,Get-OffloadDataTransferSetting,Get-OperationValidation,Get-Partition,Get-PartitionSupportedSize,Get-PcsvDevice,Get-PcsvDeviceLog,Get-PhysicalDisk,Get-PhysicalDiskStorageNodeView,Get-PhysicalExtent,Get-PhysicalExtentAssociation,Get-PnpDevice,Get-PnpDeviceProperty,Get-PrintConfiguration,Get-Printer,Get-PrinterDriver,Get-PrinterPort,Get-PrinterProperty,Get-PrintJob,Get-PSRepository,Get-ResiliencySetting,Get-ScheduledTask,Get-ScheduledTaskInfo,Get-SmbBandWidthLimit,Get-SmbClientConfiguration,Get-SmbClientNetworkInterface,Get-SmbConnection,Get-SmbDelegation,Get-SmbGlobalMapping,Get-SmbMapping,Get-SmbMultichannelConnection,Get-SmbMultichannelConstraint,Get-SmbOpenFile,Get-SmbServerConfiguration,Get-SmbServerNetworkInterface,Get-SmbSession,Get-SmbShare,Get-SmbShareAccess,Get-SmbWitnessClient,Get-StartApps,Get-StorageAdvancedProperty,Get-StorageDiagnosticInfo,Get-StorageEnclosure,Get-StorageEnclosureStorageNodeView,Get-StorageEnclosureVendorData,Get-StorageExtendedStatus,Get-StorageFaultDomain,Get-StorageFileServer,Get-StorageFirmwareInformation,Get-StorageHealthAction,Get-StorageHealthReport,Get-StorageHealthSetting,Get-StorageJob,Get-StorageNode,Get-StoragePool,Get-StorageProvider,Get-StorageReliabilityCounter,Get-StorageSetting,Get-StorageSubSystem,Get-StorageTier,Get-StorageTierSupportedSize,Get-SupportedClusterSizes,Get-SupportedFileSystems,Get-TargetPort,Get-TargetPortal,Get-TestDriveItem,Get-Verb,Get-VirtualDisk,Get-VirtualDiskSupportedSize,Get-Volume,Get-VolumeCorruptionCount,Get-VolumeScrubPolicy,Get-VpnConnection,Get-VpnConnectionTrigger,Get-WdacBidTrace,Get-WindowsUpdateLog,Get-WUAVersion,Get-WUIsPendingReboot,Get-WULastInstallationDate,Get-WULastScanSuccessDate,Grant-FileShareAccess,Grant-SmbShareAccess,help,Hide-VirtualDisk,Import-IseSnippet,Import-PowerShellDataFile,ImportSystemModules,In,Initialize-Disk,InModuleScope,Install-Dtc,Install-Module,Install-Script,Install-WUUpdates,Invoke-AsWorkflow,Invoke-Mock,Invoke-OperationValidation,Invoke-Pester,It,Lock-BitLocker,mkdir,Mock,more,Mount-DiskImage,Move-SmbWitnessClient,New-AutologgerConfig,New-DAEntryPointTableItem,New-DscChecksum,New-EapConfiguration,New-EtwTraceSession,New-FileShare,New-Fixture,New-Guid,New-IscsiTargetPortal,New-IseSnippet,New-MaskingSet,New-NetAdapterAdvancedProperty,New-NetEventSession,New-NetFirewallRule,New-NetIPAddress,New-NetIPHttpsConfiguration,New-NetIPseospSetting,New-NetIPsecMainModeCryptoSet,New-NetIPsecMainModeRule,New-NetIPsecPhase1AuthSet,New-NetIPsecPhase2AuthSet,New-NetIPsecQuickModeCryptoSet,New-NetIPsecRule,New-NetLbfoTeam,New-NetNat,New-NetNatTransitionConfiguration,New-NetNeighbor,New-NetQosPolicy,New-NetRoute,New-NetSwitchTeam,New-NetTransportFilter,New-NetworkSwitchVlan,New-Partition,New-PesterOption,New-PSWorkflowSession,New-ScheduledTask,New-ScheduledTaskAction,New-ScheduledTaskPrincipal,New-ScheduledTaskSettingsSet,New-ScheduledTaskTrigger,New-ScriptFileInfo,New-SmbGlobalMapping,New-SmbMapping,New-SmbMultichannelConstraint,New-SmbShare,New-StorageFileServer,New-StoragePool,New-StorageSubsystemVirtualDisk,New-StorageTier,New-TemporaryFile,New-VirtualDisk,New-VirtualDiskClone,New-VirtualDiskSnapshot,New-Volume,New-VpnServerAddress,Open-NetGPO,Optimize-StoragePool,Optimize-Volume,oss,Pause,prompt,PSConsoleHostReadline,Publish-Module,Publish-Script,Read-PrinterNfcTag,Register-ClusteredScheduledTask,Register-DnsClient,Register-IscsiSession,Register-PSRepository,Register-ScheduledTask,Register-StorageSubsystem,Remove-AutologgerConfig,Remove-BitLockerKeyProtector,Remove-DAEntryPointTableItem,Remove-DnsClientNrptRule,Remove-DscConfigurationDocument,Remove-DtcClusterTMMapping,Remove-EtwTraceProvider,Remove-FileShare,Remove-InitiatorId,Remove-InitiatorIdFromMaskingSet,Remove-IscsiTargetPortal,Remove-MaskingSet,Remove-MpPreference,Remove-MpThreat,Remove-NetAdapterAdvancedProperty,Remove-NetEventNetworkAdapter,Remove-NetEventPacketCaptureProvider,Remove-NetEventProvider,Remove-NetEventSession,Remove-NetEventVFPProvider,Remove-NetEventVmNetworkAdapter,Remove-NetEventVmSwitch,Remove-NetEventVmSwitchProvider,Remove-NetEventWFPCaptureProvider,Remove-NetFirewallRule,Remove-NetIPAddress,Remove-NetIPHttpsCertBinding,Remove-NetIPHttpsConfiguration,Remove-NetIPseospSetting,Remove-NetIPsecMainModeCryptoSet,Remove-NetIPsecMainModeRule,Remove-NetIPsecMainModeSA,Remove-NetIPsecPhase1AuthSet,Remove-NetIPsecPhase2AuthSet,Remove-NetIPsecQuickModeCryptoSet,Remove-NetIPsecQuickModeSA,Remove-NetIPsecRule,Remove-NetLbfoTeam,Remove-NetLbfoTeamMember,Remove-NetLbfoTeamNic,Remove-NetNat,Remove-NetNatExternalAddress,Remove-NetNatStaticMapping,Remove-NetNatTransitionConfiguration,Remove-NetNeighbor,Remove-NetQosPolicy,Remove-NetRoute,Remove-NetSwitchTeam,Remove-NetSwitchTeamMember,Remove-NetTransportFilter,Remove-NetworkSwitchEthernetPortIPAddress,Remove-NetworkSwitchVlan,Remove-Odbsn,Remove-Partition,Remove-PartitionAccessPath,Remove-PhysicalDisk,Remove-Printer,Remove-PrinterDriver,Remove-PrinterPort,Remove-PrintJob,Remove-SmbBandwidthLimit,Remove-SmbGlobalMapping,Remove-SmbMapping,Remove-SmbMultichannelConstraint,Remove-SmbShare,Remove-StorageFaultDomain,Remove-StorageFileServer,Remove-StorageHealthIntent,Remove-StorageHealthSetting,Remove-StoragePool,Remove-StorageTier,Remove-TargetPortFromMaskingSet,Remove-VirtualDisk,Remove-VirtualDiskFromMaskingSet,Remove-VpnConnection,Remove-VpnConnectionRoute,Remove-VpnConnectionTriggerApplication,Remove-VpnConnectionTriggerDnsConfiguration,Remove-VpnConnectionTriggerTrustedNetwork,Rename-DAEntryPointTableItem,Rename-MaskingSet,Rename-NetAdapter,Rename-NetFirewallRule,Rename-NetIPHttpsConfiguration,Rename-NetIPsecMainModeCryptoSet,Rename-NetIPsecMainModeRule,Rename-NetIPsecPhase1AuthSet,Rename-NetIPsecPhase2AuthSet,Rename-NetIPsecQuickModeCryptoSet,Rename-NetIPsecRule,Rename-NetLbfoTeam,Rename-NetSwitchTeam,Rename-Printer,Repair-FileIntegrity,Repair-VirtualDisk,Repair-Volume,Reset-DAClientExperienceConfiguration,Reset-DAEntryPointTableItem,Reset-DtcLog,Reset-NCSIPolicyConfiguration,Reset-Net6to4Configuration,Reset-NetAdapterAdvancedProperty,Reset-NetDnsTransitionConfiguration,Reset-NetIPHttpsConfiguration,Reset-NetIsatapConfiguration,Reset-NetTeredoConfiguration,Reset-PhysicalDisk,Reset-StorageReliabilityCounter,Resize-Partition,Resize-StorageTier,Resize-VirtualDisk,Restart-NetAdapter,Restart-PcsvDevice,Restart-PrintJob,Restore-DscConfiguration,Restore-NetworkSwitchConfiguration,Resume-BitLocker,Resume-PrintJob,Revoke-FileShareAccess,Revoke-SmbShareAccess,SafeGetCommand,Save-EtwTraceSession,Save-Module,Save-NetGPO,Save-NetworkSwitchConfiguration,Save-Script,Send-EtwTraceSession,Set-AutologgerConfig,Set-ClusteredScheduledTask,Set-DAClientExperienceConfiguration,Set-DAEntryPointTableItem,Set-Disk,Set-DnsClient,Set-DnsClientGlobalSetting,Set-DnsClientNrptGlobal,Set-DnsClientNrptRule,Set-DnsClientServerAddress,Set-DtcAdvancedHostSetting,Set-DtcAdvancedSetting,Set-DtcClusterDefault,Set-DtcClusterTMMapping,Set-Dtefault,Set-DtcLog,Set-DtcNetworkSetting,Set-DtcTransaction,Set-DtcTransactionsTraceSession,Set-DtcTransactionsTraceSetting,Set-DynamicParameterVariables,Set-EtwTraceProvider,Set-FileIntegrity,Set-FileShare,Set-FileStorageTier,Set-InitiatorPort,Set-IscsiChapSecret,Set-LogProperties,Set-MMAgent,Set-MpPreference,Set-NCSIPolicyConfiguration,Set-Net6to4Configuration,Set-NetAdapter,Set-NetAdapterAdvancedProperty,Set-NetAdapterBinding,Set-NetAdapterChecksumOffload,Set-NetAdapterEncapsulatedPacketTaskOffload,Set-NetAdapterIPsecOffload,Set-NetAdapterLso,Set-NetAdapterPacketDirect,Set-NetAdapterPowerManagement,Set-NetAdapterQos,Set-NetAdapterRdma,Set-NetAdapterRsc,Set-NetAdapterRss,Set-NetAdapterSriov,Set-NetAdapterVmq,Set-NetConnectionProfile,Set-NetDnsTransitionConfiguration,Set-NetEventPacketCaptureProvider,Set-NetEventProvider,Set-NetEventSession,Set-NetEventVFPProvider,Set-NetEventVmSwitchProvider,Set-NetEventWFPCaptureProvider,Set-NetFirewallAddressFilter,Set-NetFirewallApplicationFilter,Set-NetFirewallInterfaceFilter,Set-NetFirewallInterfaceTypeFilter,Set-NetFirewallPortFilter,Set-NetFirewallProfile,Set-NetFirewallRule,Set-NetFirewallSecurityFilter,Set-NetFirewallServiceFilter,Set-NetFirewallSetting,Set-NetIPAddress,Set-NetIPHttpsConfiguration,Set-NetIPInterface,Set-NetIPseospSetting,Set-NetIPsecMainModeCryptoSet,Set-NetIPsecMainModeRule,Set-NetIPsecPhase1AuthSet,Set-NetIPsecPhase2AuthSet,Set-NetIPsecQuickModeCryptoSet,Set-NetIPsecRule,Set-NetIPv4Protocol,Set-NetIPv6Protocol,Set-NetIsatapConfiguration,Set-NetLbfoTeam,Set-NetLbfoTeamMember,Set-NetLbfoTeamNic,Set-NetNat,Set-NetNatGlobal,Set-NetNatTransitionConfiguration,Set-NetNeighbor,Set-NetOffloadGlobalSetting,Set-NetQosPolicy,Set-NetRoute,Set-NetTCPSetting,Set-NetTeredoConfiguration,Set-NetUDPSetting,Set-NetworkSwitchEthernetPortIPAddress,Set-NetworkSwitchPortMode,Set-NetworkSwitchPortProperty,Set-NetworkSwitchVlanProperty,Set-Odbriver,Set-Odbsn,Set-Partition,Set-PcsvDeviceBootConfiguration,Set-PcsvDeviceNetworkConfiguration,Set-PcsvDeviceUserPassword,Set-PhysicalDisk,Set-PrintConfiguration,Set-Printer,Set-PrinterProperty,Set-PSRepository,Set-ResiliencySetting,Set-ScheduledTask,Set-SmbBandwidthLimit,Set-SmbClientConfiguration,Set-SmbPathAcl,Set-SmbServerConfiguration,Set-SmbShare,Set-StorageFileServer,Set-StorageHealthSetting,Set-StoragePool,Set-StorageProvider,Set-StorageSetting,Set-StorageSubSystem,Set-StorageTier,Set-TestInconclusive,Setup,Set-VirtualDisk,Set-Volume,Set-VolumeScrubPolicy,Set-VpnConnection,Set-VpnConnectionIPsecConfiguration,Set-VpnConnectionProxy,Set-VpnConnectionTriggerDnsConfiguration,Set-VpnConnectionTriggerTrustedNetwork,Should,Show-NetFirewallRule,Show-NetIPsecRule,Show-VirtualDisk,Start-AppBackgroundTask,Start-AutologgerConfig,Start-Dtc,Start-DtcTransactionsTraceSession,Start-EtwTraceSession,Start-MpScan,Start-MpWDOScan,Start-NetEventSession,Start-PcsvDevice,Start-ScheduledTask,Start-StorageDiagnosticLog,Start-Trace,Start-WUScan,Stop-DscConfiguration,Stop-Dtc,Stop-DtcTransactionsTraceSession,Stop-EtwTraceSession,Stop-NetEventSession,Stop-PcsvDevice,Stop-ScheduledTask,Stop-StorageDiagnosticLog,Stop-StorageJob,Stop-Trace,Suspend-BitLocker,Suspend-PrintJob,Sync-NetIPsecRule,TabExpansion2,Test-Dtc,Test-NetConnection,Test-ScriptFileInfo,Unblock-FileShareAccess,Unblock-SmbShareAccess,Uninstall-Dtc,Uninstall-Module,Uninstall-Script,Unlock-BitLocker,Unregister-AppBackgroundTask,Unregister-ClusteredScheduledTask,Unregister-IscsiSession,Unregister-PSRepository,Unregister-ScheduledTask,Unregister-StorageSubsystem,Update-Disk,Update-DscConfiguration,Update-EtwTraceSession,Update-HostStorageCache,Update-IscsiTarget,Update-IscsiTargetPortal,Update-Module,Update-ModuleManifest,Update-MpSignature,Update-NetIPsecRule,Update-Script,Update-ScriptFileInfo,Update-SmbMultichannelConnection,Update-StorageFirmware,Update-StoragePool,Update-StorageProviderCache,Write-DtcTransactionsTraceSession,Write-PrinterNfcTag,Write-VolumeCache
	},
	morekeywords={Do,Else,For,ForEach,Function,If,In,Until,While},
	alsodigit={-},
	sensitive=false,
	morecomment=[l]{\#},
	morecomment=[n]{<\#}{\#>},
	morestring=[b]{"},
	morestring=[b]{'},
	morestring=[s]{@'}{'@},
	morestring=[s]{@"}{"@}
}
\usepackage{pifont}
\usepackage[inline]{enumitem}
\definecolor{my_orange}{HTML}{F28522}
\definecolor{my_red}{HTML}{FF1F5B}
\definecolor{my_green}{HTML}{00CD6C}

\newcommand{\tick}{\textcolor{my_green}{\ding{51}}} 
\newcommand{\cross}{\textcolor{my_red}{\ding{55}}}  

\newcommand{\ColorVT}[2]{%
    \IfStrEq{#2}{fp}{\cellcolor{lightred}#1}{%
        \IfStrEq{#2}{not found}{\cellcolor{lightred}#1}{#1}%
    }%
}

\newcommand{\ColorFP}[1]{%
    \ifnum\pdfstrcmp{#1}{fp}=0 \textcolor{my_orange}{#1}%
    \else #1%
    \fi
}

\newcommand{\MapFound}[1]{%
    \ifnum\pdfstrcmp{#1}{found}=0 \tick%
    \else\ifnum\pdfstrcmp{#1}{not found}=0 \cross%
    \else \ColorFP{#1}%
    \fi\fi
}

\newcommand{\MapCapa}[1]{%
    \ifnum\pdfstrcmp{#1}{yes}=0 \tick%
    \else\ifnum\pdfstrcmp{#1}{no}=0 \cross%
    \else #1%
    \fi\fi
}

\newcommand{\ColorCell}[1]{%
    \ifnum\pdfstrcmp{#1}{fp}=0%
        \cellcolor{yellow}{#1}%
    \else%
        \ifnum\pdfstrcmp{#1}{found}=0%
            \cellcolor{green}{#1}%
        \else%
            \ifnum\pdfstrcmp{#1}{not found}=0%
                \cellcolor{red}{#1}%
            \else%
                #1%
            \fi%
        \fi%
    \fi%
}

\definecolor{bgcolor}{rgb}{0.95,0.95,0.95}

\usepgfplotslibrary{colorbrewer}
\pgfplotsset{cycle list/Dark2-8}
\pgfplotsset{compat = 1.18} 
\usetikzlibrary{pgfplots.statistics} 

\definecolor{whitesmoke}{rgb}{0.96, 0.96, 0.96}
\definecolor{floralwhite}{rgb}{1.0, 0.98, 0.94}

\title{Coding Malware in Fancy Programming Languages\\ for Fun and Profit}

\author[1]{Theodoros Apostolopoulos}
\author[1]{Vasilios Koutsokostas}
\author[1]{Nikolaos Totosis}
\author[1,2]{Constantinos Patsakis}
\author[3]{Georgios Smaragdakis}
\affil[1]{Department of Informatics, University of Piraeus, 80 Karaoli \& Dimitriou str., 18534 Piraeus, Greece}

\affil[2]{Information Management Systems Institute of Athena Research Centre, Greece}
\affil[3]{Delft University of Technology, Netherlands}

\begin{document}

\date{}
\maketitle
\begin{abstract} 
    The continuous increase in malware samples, both in sophistication and number, presents many challenges for organizations and analysts, who must cope with thousands of new heterogeneous samples daily. This requires robust methods to quickly determine whether a file is malicious. Due to its speed and efficiency, static analysis is the first line of defense. 
    
    In this work, we illustrate how the practical state-of-the-art methods used by antivirus solutions may fail to detect evident malware traces. The reason is that they highly depend on very strict signatures where minor deviations prevent them from detecting shellcodes that otherwise would immediately be flagged as malicious. Thus, our findings illustrate that malware authors may drastically decrease the detections by converting the code base to less-used programming languages. To this end, we study the features that such programming languages introduce in executables and the practical issues that arise for practitioners to detect malicious activity.
\end{abstract}

\section{Introduction}

In the past decade, malware has undergone significant changes. The main drivers of these changes can be attributed to the vast digitization of products and services and the development of a payment system that allows anonymous transactions to bypass the protections of the traditional banking system. The former has boosted the number of possible victims and the potential impact of malware. Moreover, anonymous payment methods enable a wide array of illicit transactions to be performed, which, in the case of malware, is the apparent case of ransomware. 
Both the US Cybersecurity and
Infrastructure Security Agency (CISA)~\cite{circia,CISA-cost-incident} and the European Union Agency for Cybersecurity
(ENISA)~\cite{ENISA-threat2023} have recognized malware as the top cyber threat. Indeed, malware attacks impact our everyday lives by harvesting sensitive information, crippling critical services, and causing significant damage to individuals and corporations~\cite{CISA-cost-incident}. This has placed malware in a pivotal role in the crime ecosystem and created an individual ecosystem with independent roles operating in a business model called Malware-as-a-Service~\cite{maas}.

The security industry's response to the abovementioned threats is collecting and analyzing malware samples. At a rate of around 280,000 malware samples per day in 2024~\cite{avatlas}, which is more or less similar to previous years, static analysis remains the most effective and profound remedy to detect malicious files quickly. In this arms race between malicious actors and defenders, the development of malware has evolved into an underground industry to bypass security controls by employing malware authors and monetizing the infected hosts. Of course, bypassing static analysis does not grant them a foothold to the targeted host. Nevertheless, it significantly raises their chances of achieving their goal, as they often need to bypass behavioral checks. Although endpoint detection and response systems usually apply such checks, and vendors often portray them as silver bullets, there are several ways to bypass them~\cite{karantzas2021empirical}. In this work, we limit our scope to static analysis.  

Even though malware written in C continues to be the most prevalent (see our analysis in Section~\ref{motivation}), malware operators, primarily known threat groups such as APT29~\cite{APT29}, increasingly include non-typical malware programming languages in their arsenal. For instance, APT29 recently used Python in their Masepie malware against Ukraine~\cite{cert_ua}, while in their Zebrocy malware, they used a mixture of Delphi, Python, C\#, and Go~\cite{salad}. Likewise, Akira ransomware shifted from C++ to Rust~\cite{akira}, BlackByte ransomware shifted from C\# to Go~\cite{Zscaler}, and Hive was ported to Rust~\cite{newhive}. According to the reports, the above changes exhibited increased resistance to reverse engineering and a low detection rate or misclassification. 

On other occasions, C-language malware families are not recreated from scratch. Instead, malware authors write loaders, droppers, and wrappers in {\it "exotic"} languages. This provides them with several advantages, e.g., bypassing signature-based detection, so they can effectively place their payloads in harder-to-detect shells that are newly built. Thus, attackers continue to use the same initial penetration vector and a significant portion of their methods, suggesting that threat actors prefer to transfer the original malware code to different languages instead of modifying their tactics, techniques, and procedures (TTPs) to avoid detection. This approach allows them to maintain the effectiveness of their attacks while remaining under the radar of security systems. Since these languages may be less widely recognized or understood, they add an extra layer of obfuscation to malware, making it harder to detect and analyze. Furthermore, security analysts have reported increased difficulty in reverse engineering such malware samples due to reprogramming efforts~\cite{uncommon-languages}. Thus, combining different languages and obfuscation techniques complicates dissecting the malware's structure, functionality, and intent.

Our work explores the problem of detecting malware written in uncommon languages using a data-driven approach.   
Rather than merely reporting and examining this trend, we performed a targeted experiment by writing malicious samples in different programming languages and compilers and drilling down to the distinctive characteristics. This analysis practically shows the unique features that adversaries gain and highlights the emerging issues for malware detection and analysis. 

The above leads to the formulation of some interesting research questions that have not been systematically studied in the academic literature, and we try to answer them in this work: 
\begin{enumerate}
\item [\textbf{RQ1}:] How does the programming language and compiler choice impact the malware detection rate?
\item [\textbf{RQ2}:] What is the root cause of this disparity?
\item [\textbf{RQ3}:] What are the benefits of an attacker shifting the codebase to less common pairs of programming language and compiler beyond the detection rate by static analysis?
\end{enumerate}

The remainder of this article is structured as follows. In the following section, we provide an overview of the related work. Next, we detail our motivation, formalize our research questions, and define our methodology. Then, in Section 4, we present our experiments and report our findings. The latter led us to examine the intrinsic differences when reverse engineering a binary in a non-standard programming language. We discuss our findings in Section 5, and finally, the work concludes, summarizing our findings and contributions and proposing ideas for future research.   

\section{Related work}
Previous but sparse research has demonstrated how the runtime mechanics of programming languages or compiler characteristics have been exploited to evade static analysis and hinder reverse engineering. For example, Wang et al.~\cite{7467351} introduced the concept of translingual obfuscation, which leverages the unique features of logic programming languages like Prolog to obscure both data layout and control flow of C programs, complicating reverse engineering efforts. Their tool, BABEL, translates C functions into Prolog predicates, leveraging Prolog's unification and backtracking to create obfuscation layers resistant to static and dynamic analysis. In~\cite{Lu_2019}, binary obfuscation has been achieved using Continuation-Passing Style (CPS) transformation, which shares similar ideas with the intermediate code produced by functional language compilers such as Haskell. CPS transformation converts control flow into continuations, which severely fragment control flow graphs (CFGs), making static analysis and reverse engineering significantly more complex. Similarly, Lambda Obfuscation proposed by Lan et al.~\cite{e1c0deb42df449bd9dc9576b1647e3a4} uses lambda calculus to obfuscate program control flow and conceal sensitive branch conditions. Replacing conditional instructions with lambda expressions prevents adversaries from leveraging symbolic execution tools to recover a program's internal logic, hindering reverse engineering efforts. In a similar train of thought, Wang et al.~\cite{wang2018turing} try to obfuscate the program execution flow by simulating Turing machines under branch conditions. Pawlowski et al.~\cite{pawlowski2016probfuscation} obfuscate the control flow by making the execution probabilistic so that the execution traces differ per execution, even on the same input, confusing this way the analyst.

Romano et al.~\cite{9833626} partially translated parts of JavaScript to Web\-Assembly to make their malware evasive. Koutsokostas and Patsakis~\cite{koutsokostas_python_2021} introduced another evasion method focusing on Python malware packaged using PyInstaller. They illustrated how AV systems inherently struggle to detect Python bytecode, allowing malware authors to evade static analysis by exploiting the packaging noise PyInstaller adds to executables. In another study~\cite{casolare2faces2022}, Casolare et al. explored malware models that exploit the dynamic features of languages like Java. By leveraging Java's dynamic compilation, reflection, and class loading mechanisms, they managed to escape device antimalware software and signature detectors. Finally, in ~\cite{10226157}, a comparative analysis of real-world malware written in C and Rust is presented, and a dedicated framework that can easily analyze Rust malware is proposed, stressing the lack of academic research on Rust malware.

For years, ransomware groups have been switching to newer, unconventional languages to make reverse engineering and detection more difficult. Moreover, various threat actors have used this approach, employing a wide range of programming languages and techniques to obfuscate their malicious code.
In~\cite{hated}, Visual Basic 6 binaries were characterized as the "\textit{most hated binaries}" among security researchers due to the complexity of reverse engineering the code to analyze malware as the tools to dissect such binaries were scarce at that time. Visual Basic 6, despite being an older language, introduced unique challenges in dissecting the malware's structure and functionality, hindering the efforts of researchers to understand and mitigate threats. The Flame malware, discovered in 2012, was dubbed "\textit{the most complex malware ever found}"~\cite{flame} at that time. It used the Lua scripting language, which was relatively uncommon in malware at that time. Incorporating Lua added a layer of obfuscation, making the malware more challenging to analyze and understand. The Duqu malware~\cite{Duqu}, also known as Stuxnet 2.0, was written primarily in C++. However, the unique assembly patterns observed in the compiled code initially led researchers to believe that it was written in an unknown high-level object-oriented programming language. After seeking help from the community~\cite{unknown}, Kaspersky Lab discovered that the unusual patterns were due to an old C++ compiler used in legacy IBM systems, which had generated the code. This revelation highlighted the challenges researchers face in understanding this complex malware. A virus called Grip contained a Brainfuck interpreter coded in Assembly to generate its keycodes. Brainfuck is an intentionally minimalistic and challenging-to-understand programming language. Another extreme example is presented in~\cite{rebol}, where attackers leverage REBOL, a lightweight language, to establish a command-and-control environment, allowing them to execute commands remotely. By employing such obscure languages, threat actors further hindered the efforts of security researchers to analyze and reverse engineer their malicious code.

To obscure the first step of the infection process and avoid security measures that identify the most common types of malicious code, malware authors can simply "wrap" commodity malware in loaders and droppers written in exotic languages. Also, malware developers can completely rewrite the code of current malware to produce new varieties. For example, the RustyBuer~\cite{RustyBuer} malware variant is a new form of the Buer malware loader. Both of these tactics are being abused by known threat actors. The Sednit group – also known as APT28~\cite{cert_ua}, Fancy Bear, Sofacy, and STRONTIUM, are among the groups that have adopted a multi-language kill chain with uncommon languages in its development process on several occasions. For instance, APT28 developed the Zebrocy backdoor in Go and then rewrote its downloader in Nim in 2019 after it was initially created in Delphi. APT28 continues to employ the same initial penetration vector and many of the same methods, implying that threat actors are more likely to change the original malware code to a different language rather than change their TTPs to avoid detection. Recently, the Tomiris APT group was spotted utilizing a polyglot arsenal of programming languages, including some uncommon or unconventional in malware development. This diversification approach appears to aim at equipping operators with ``full-spectrum malware'' capable of evading security products. In several observed instances, the actor persistently cycled through different language malware strains until one was successfully executed on the targeted machines~\cite{tomiris}.

Exotic programming languages provide extra levels of obfuscation that go beyond traditional security procedures. Also, these languages are less often used in malware, reverse engineers are less experienced with their implementation, and malware analysis tools and sandboxes have a hard time evaluating samples written in them. Malware rewrites disrupt the static signature produced for well-known malware families, and because there is no identifying signature, malware written in obscure languages frequently escapes unnoticed by antivirus software.
Malware detection using signatures relies on the presence of specific static characteristics within a file that remain constant and do not require execution to be identified. When malware is built in a new language, static indicators (for example, YARA rules) become irrelevant or ineffective~\cite{enisa}.

Malware samples written in uncommon languages can multiply the effort required for reverse engineering by a sufficient factor. Many of these languages, particularly functional ones (e.g., Haskell, Lisp), employ a vastly different execution model from traditional malware development languages like C. In addition, these languages often introduce a large number of functions to the executable as part of their standard environment, resulting in a bloated binary that makes even simple programs like "Hello world" contain thousands of functions (e.g., Dart and Go). Moreover, using unconventional programming languages also introduces additional challenges to analysts, such as indirect function calls, different evaluation models, error handling procedures, memory safety operations, and garbage collectors. They also contain unique data structures and calling conventions, "mangled" symbols, as well as unique stack and heap management systems. Specifically, functional languages are characterized by their use of immutable data structures, first-class functions, and lazy evaluation, which can result in code that is difficult to comprehend and reverse engineer. In addition, different compilation options or compiler versions can make analysis even more challenging by breaking usual reverse engineering patterns. Furthermore, each language's macro and meta-programming capabilities can help to further obfuscate the binary and slow down the analysis. The combination of the aforementioned artifacts can easily confuse the malware analyst and state-of-the-art tools, leading them to an unproductive rabbit hole. According to~\cite{298910}, many security experts consider that alternative languages like Golang, Rust, and Delphi produce compiled programs that are significantly less straightforward to analyze compared to traditional C-based binaries. In fact, as stated, many consider using such languages to be a novel
evasive technique and the lack of tools to deal with a rising problem, as existing tools may produce less accurate results. Also, recent actions like the project OxA11C~\cite{s1} launched by Sentinel One and Intezer Team, which aims to develop a methodology to make reverse engineering of Rust malware more approachable, as well as develop new tools to help researchers showcase the extent of the problem in the malware analysis domain.

\section{Motivation and Methodology}\label{motivation}

\begin{figure}[t]
    \centering
    \includegraphics[width=.8\linewidth]{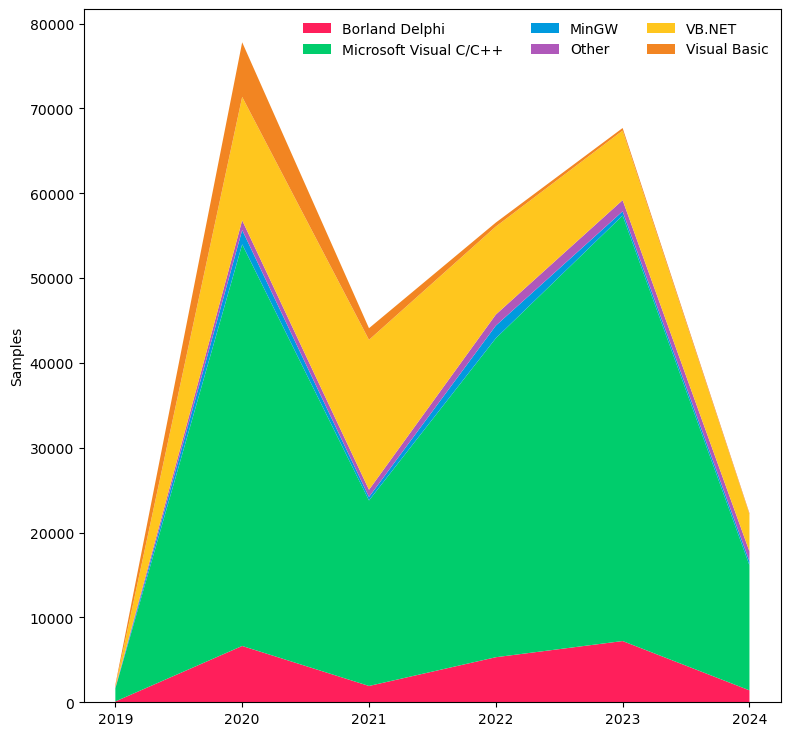}
        \vspace{-1em}
    \caption{Distribution of the top 5 programming languages of samples per year.}
        \vspace{-1em}
    \label{fig:distribution}
\end{figure}

\begin{figure*}[!th]
    \centering
    \includegraphics[width=\linewidth]{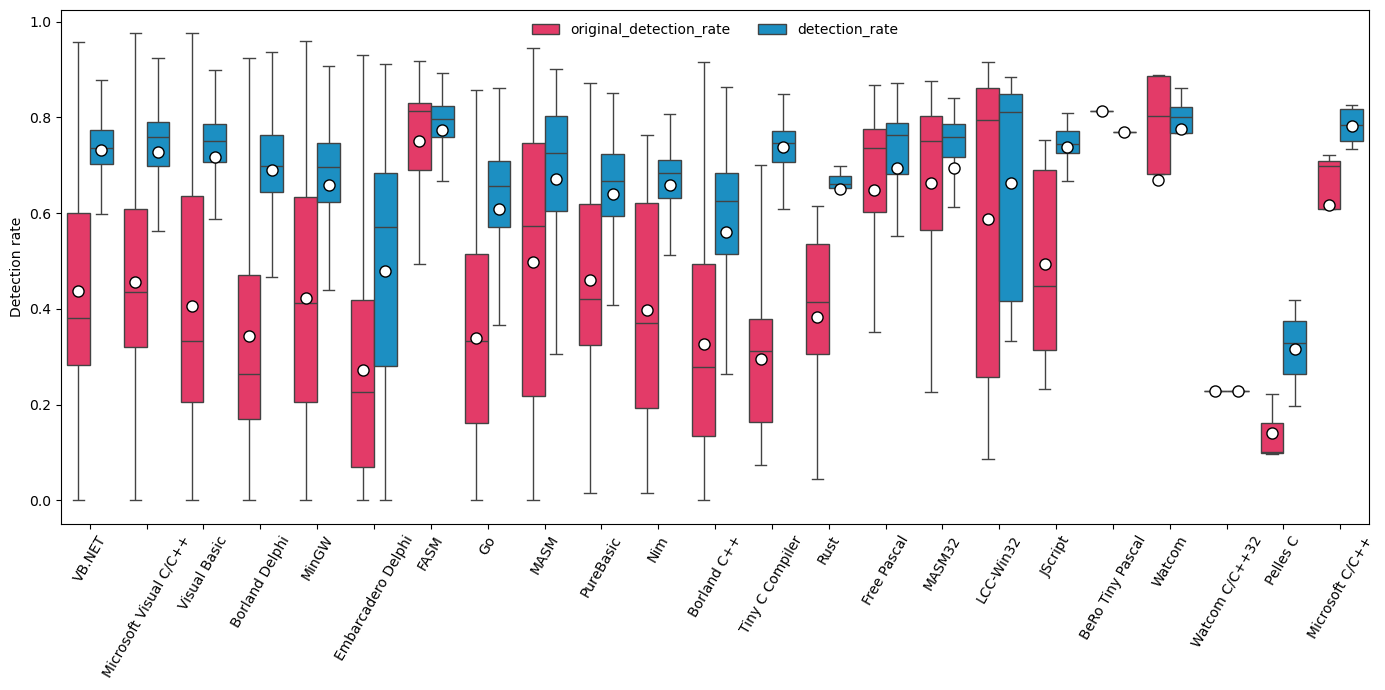} 
    \caption{Deviations in the detection rate per programming language.}
    \label{fig:detection_rate}
\end{figure*}

We used real-world public datasets to establish ground truth on the usage of various programming languages and compilers by malware authors. First, we used the latest export of the database of Malware Bazaar~\cite{abuse}. We limited the dataset to compiled files and, more precisely, Windows executables. At the time of writing, this export contains 399,043 Windows executable files, adhering to the portable executable format. In general, \texttt{EXE} files are consistently the file format with the highest number of submissions and detections~\cite{vt}. We queried these files with their hashes in VirusTotal, collecting the detection rate and, where available, the programming language and compiler. As shown in Figure \ref{fig:distribution}, there are trends in the usage of programming languages and compilers by malware authors. These trends do not follow the trends of the TIOBE index~\cite{tiobe}, e.g., Python and Java, the two programming languages most widely used, are not represented in the dataset. Nevertheless, the differences are stiffer. For instance, the deviations in the detection rate by programming language and compiler are more than apparent, Figure \ref{fig:detection_rate}. One can clearly observe the deviation in the detection rate from the first submission to the latest one. Moreover, it is also apparent that this disparity is wider in the less-used programming languages and compilers. Notably, this disparity appears in both the original and latest detections. Even more alarmingly, in this segment, we observe the lowest detection rate in the latest detections, showing that even malware in well-known languages has a low detection rate when a less common compiler is used. 

Close investigation shows that the programming language and compiler choice can significantly impact the detection rate; see Figure \ref{fig:detection_rate}. While one would expect less used programming languages, e.g., Rust and Nim, to have worse detection rates because the sparsity of samples would not allow the creation of robust rules, the use of non-widely used compilers, e.g., Pelles C, Embarcadero Delphi, and Tiny C, has a more substantial impact on the detection rate.

\begin{figure*}[!th]
    \centering
    \includegraphics[width=\linewidth]{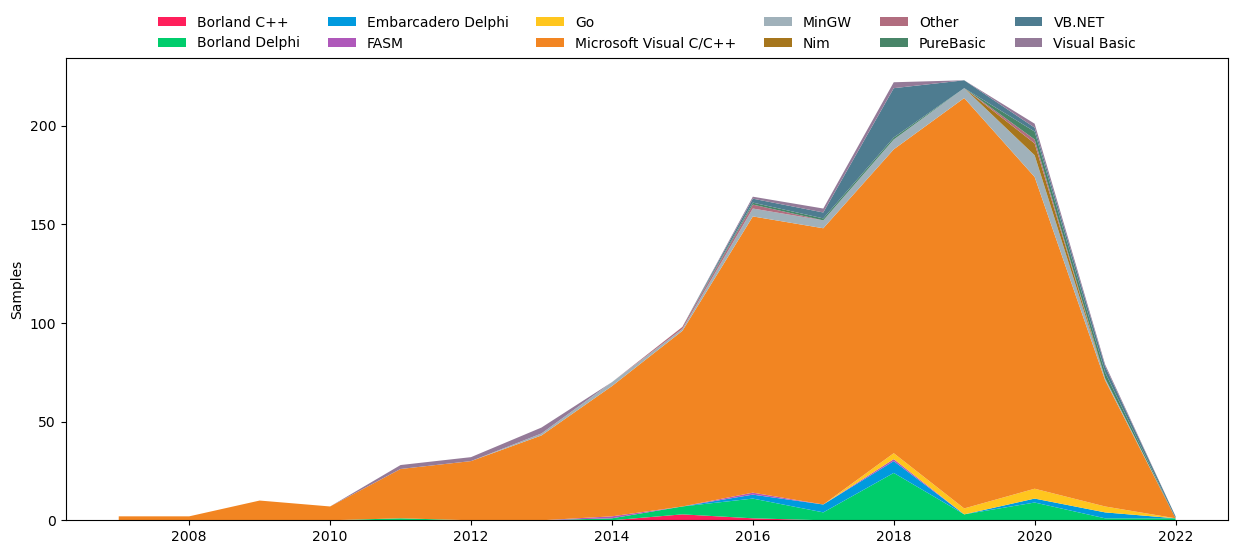} 
        \vspace{-1em}
    \caption{Distribution of top programming languages of APT samples per year from \cite{gonzalez2023technical}.}
    \label{fig:APTdistribution}
\end{figure*}

Then, we moved on to a more specific dataset. More precisely, we used the dataset of Gonz{\'a}lez-Manzano et al.~\cite{gonzalez2023technical}, which is focused on APTs, making our findings more focused. Limiting the dataset to PE executables (2,190 samples), one can clearly observe in Figure \ref{fig:APTdistribution} that the malware authors have shifted from coding \emph{only} in Microsoft C++ to using more languages and compilers. Indeed, as time goes by, APTs choose more diverse programming languages and compilers, e.g., Borland and Embarcadero Delphi, Borland and Microsoft C++, or Purebasic. Apparently, these trends are aligned with the findings of the larger Bazaar dataset.

To answer the research questions (see Section \S 1), we developed a specific methodology and performed some very targeted experiments. According to our methodology, first, we create a reference dataset with malicious binaries. The intention is to make it as heterogeneous as possible in terms of programming languages and compilers. Nevertheless, we deliberately add well-known payloads that are immediately flagged by antimalware engines and do not obfuscate the binaries. This way, we avoid possible biases that obfuscation methods can introduce. Then, we submit the binaries to VirusTotal to assess how detectable these samples are from commercial antimalware engines (RQ1). We analyze the binaries to determine their structural differences, use tools and custom scripts to quantify their differences at the binary level (RQ2), and examine the effort and drawbacks that a reverse engineer would have. The latter, along with the known advantages of some frameworks and programming languages, allow us to streamline the benefits of a malware author to shift her codebase into less-used programming languages or use less common compilers (RQ3).

\section{Experiment}

In this section, we empirically show that the challenges posed by uncommon programming languages create difficulties for malware analysts during reverse engineering and affect automated detection systems.

\subsection{Setting Up the Experiment}

We performed a scoped experiment to assess the robustness of static analysis methods against malware written in uncommon (less-used) programming languages and frameworks. To this end, we experimented with common and uncommon programming languages that can generate native standalone Windows PE files using either a compiler or a packager. We selected the languages so that they can interact with the Windows OS by exposing a system command or by interacting with the Windows API via built-in libraries or the Foreign Function Interface (FFI). From a macro perspective, interaction with the underlying OS is the bare bone of every malware, alongside networking and cryptographic functionality.

\begin{lstlisting}[language=Powershell,label={payload2},caption={Payload I - PowerShell reverse shell.}]
    powershell -NoP -NonI -W Hidden -Exec Bypass -Command New-Object System.Net.Sockets.TCPClient($IP,$port);$stream=$client.GetStream();[byte[]]$bytes=0..65535|%{0};while(($i=$stream.Read($bytes,0,$bytes.Length)) -ne 0){$data=(New-Object -TypeName System.Text.ASCIIEncoding).GetString($bytes,0,$i);$sendback=(iex $data 2>&1 | Out-String);$sendback2=$sendback + 'PS ' + (pwd).Path + '> ';$sendbyte=([text.encoding]::ASCII).GetBytes($sendback2);$stream.Write($sendbyte,0,$sendbyte.Length);$stream.Flush();};$client.Close()
\end{lstlisting}

\begin{lstlisting}[language=Powershell,label={payload2},caption={Payload II - Vanilla shellcode execution in C.}]
LPVOID addressPointer = VirtualAlloc(NULL, sizeof(shellcode), 0x3000, 0x40);
RtlMoveMemory(addressPointer, shellcode, sizeof(shellcode));
HANDLE handle = CreateThread(NULL, 0, (LPTHREAD_START_ROUTINE)addressPointer, NULL, 0, 0);
WaitForSingleObject(handle, -1);
\end{lstlisting}

Additionally, we tried to cover as many programming paradigms as feasible, as long as the produced binaries were standalone and dependent dynamically only on the native Windows DLLs or the .NET Framework. We created executables containing very simple and well-known payloads to evaluate the detection rate of malware written in different programming languages. The payloads were chosen from lists of online reports containing the most critical MITRE techniques used by adversaries~\cite{mitre15}, particularly the T1059 Command and Scripting Interpreter~\cite{T1059} and the T1055 Process Injection~\cite{T1055}. The first class of executables issues a system command that calls Powershell to initiate a reverse shell via a well-known piped command (see Listing~\ref{payload1}). In contrast, the second class executes shellcode using a standard sequence of Windows API functions (see Listing~\ref{payload2}). The actual shellcode invokes an executable and acts as a loader. To keep the analysis consistent, we tried to construct homogeneous and simple samples in terms of translation from one language to another without employing any techniques of obfuscation, anti-analysis, or compiler optimization. The code and corresponding binaries are available on GitHub~\cite{repo}.

Then, we examine the binaries from two different standpoints. First, from the static analysis perspective using state-of-the-art antivirus engines and custom and open-source static analysis tools. Second, from the perspective of automated reverse engineering.

\subsection{VirusTotal Results}

In this part of the experiment, we used 39 programming languages and 50 different compilers or packagers to generate two samples for each possible payload, producing 100 unique samples. Then, we uploaded them to VirusTotal~\cite{VirusTotal} and reported the detection results; see Table~\ref{tab:VT_results}. It should be noted that, despite many of these samples being uploaded to VirusTotal for more than a year, a surprising number of them still remains undetected to date, even after rescans.

\begin{figure}[!th]
\centering
    \begin{subfigure}[hb]{0.485\textwidth}
    	\centering
          \includegraphics[width=\textwidth]{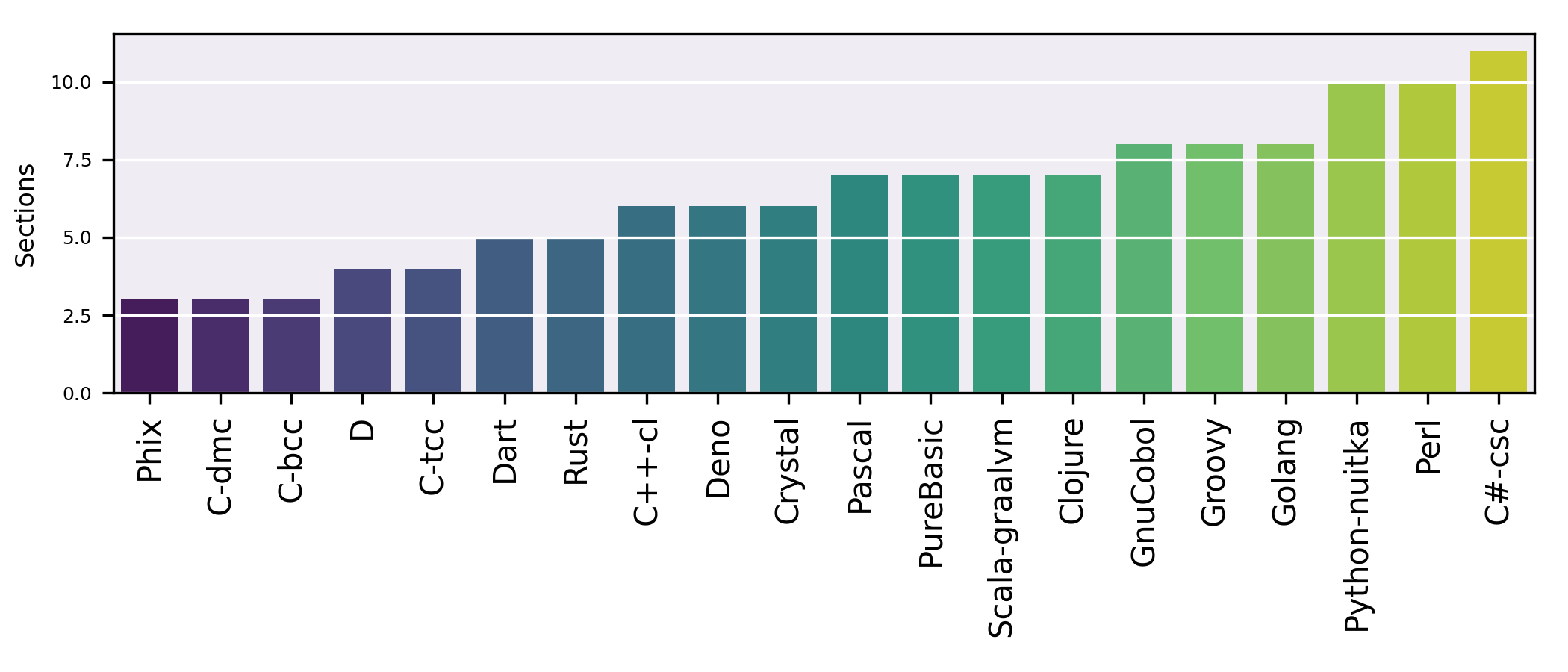}
          \caption{Variation on the number of sections per payload/language.}
          \label{fig:box_sections}
       \end{subfigure}
    \begin{subfigure}[hb]{0.485\textwidth}
    	\centering
          \includegraphics[width=\textwidth]{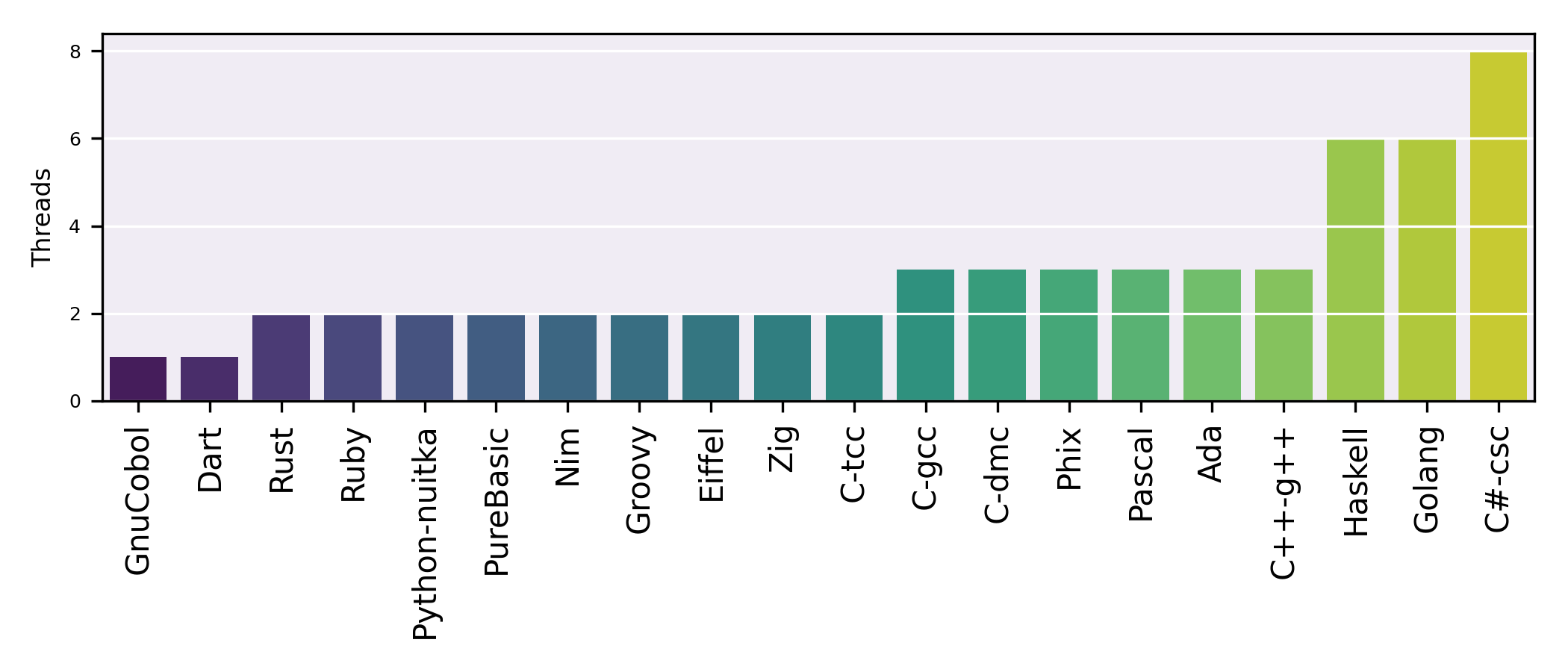}
          \caption{Variation on the number of threads per payload/language.}
          \label{fig:box_threads}
       \end{subfigure}
       
    \begin{subfigure}[hb]{0.485\textwidth}
    	\centering
          \includegraphics[width=\textwidth]{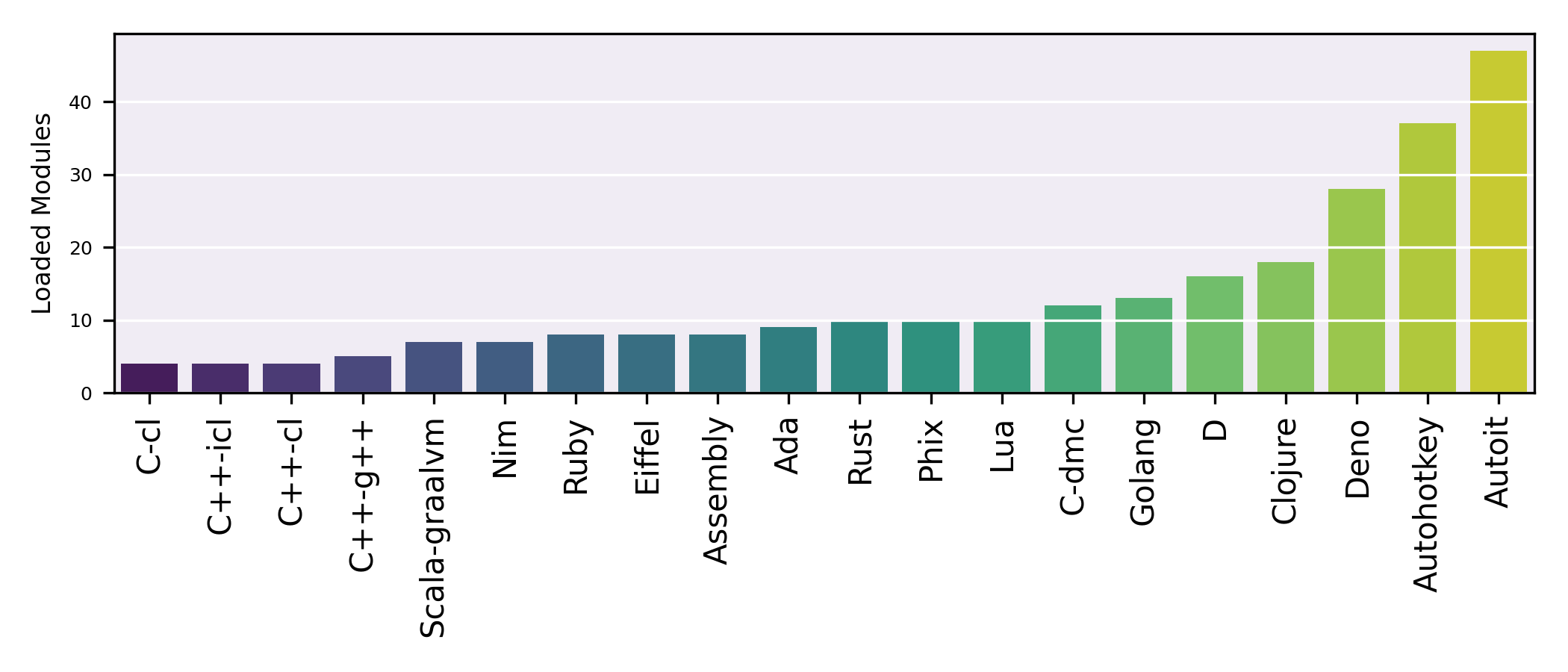}
          \caption{Variation on the number of loaded DLLs per language/language.}
          \label{fig:box_dlls}
       \end{subfigure}
        \begin{subfigure}[hb]{0.485\textwidth}
    	\centering
          \includegraphics[width=\textwidth]{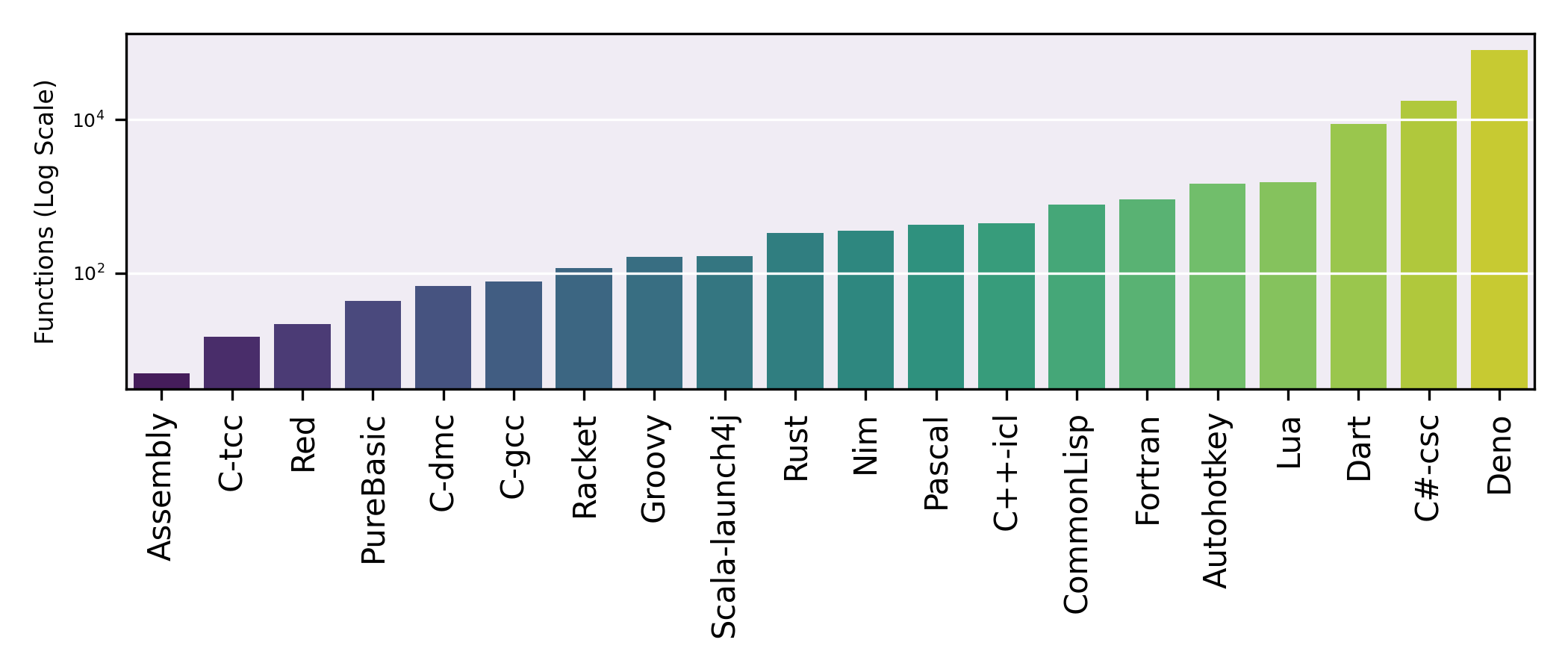}
          \caption{Variation on the number of functions per language/language.}
          \label{fig:functions}
       \end{subfigure}
       \begin{subfigure}[hb]{0.485\textwidth}
    	\centering
          \includegraphics[width=\textwidth]{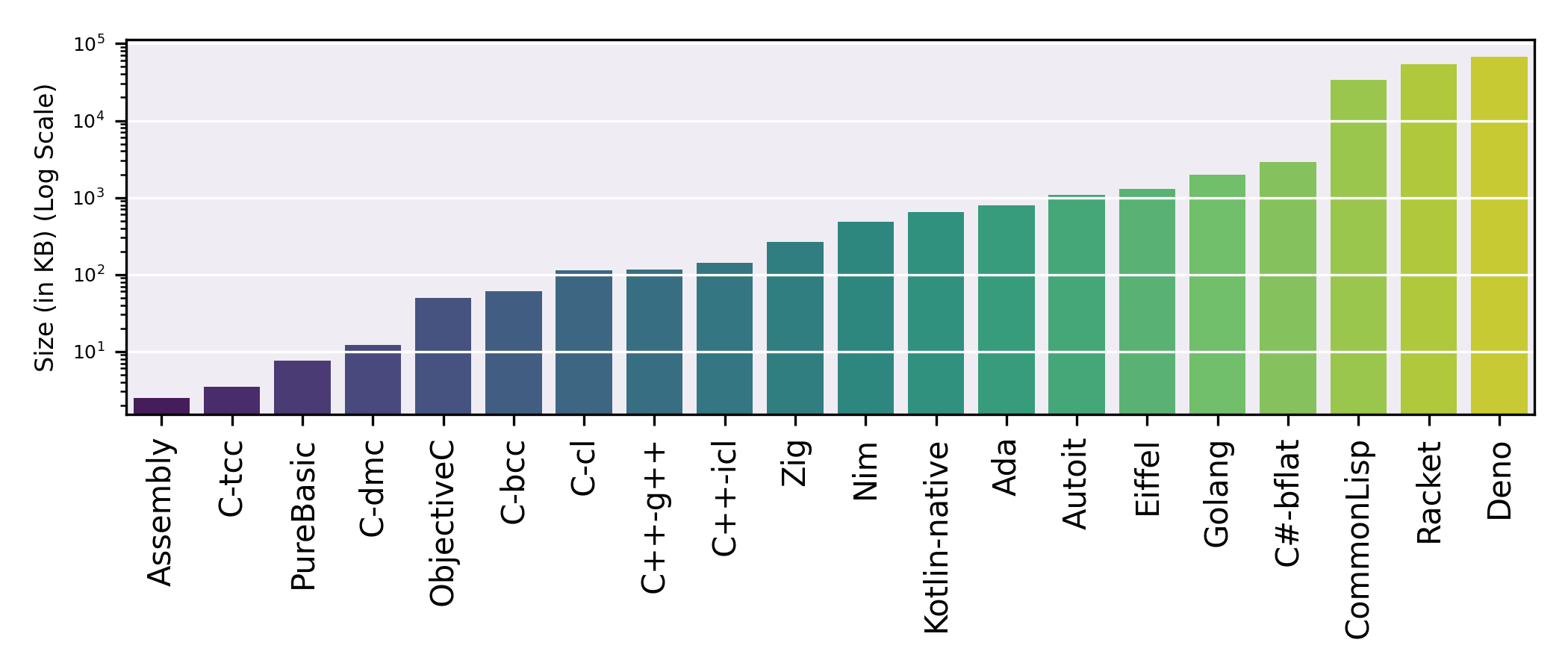}
          \caption{Variation on the size of executable per language/language.}
          \label{fig:size}
       \end{subfigure}
       \caption{Barplots to illustrate the variation of shellcode samples.}
       \label{fig:boxplots}
       
\end{figure}

One can observe a great variance in the detection rate of the same payloads not only between different programming languages but also between compilers for the same programming language (see, for example, C/gcc vs C/dmc). Quite alarmingly, for the first payload, there are 13 samples with zero detections from AVs and 19 samples that reported very low detection rate (less than 5 AV engines), meaning that 32 of the 50 samples went undetected and had an overall detection rate of 6\%. In the case of the shellcode payload, we had two samples with zero detections and 11 samples having very low detection rates with generic AV signatures like \texttt{Malicious(Moderate Confidence)}, \texttt{W64.AIDetectMalware}, \texttt{Unsafe}, \texttt{Python.Shell.6} (while not being a Python sample) and \texttt{Trojan.Malware.300983.susgen}, a notorious false positive detection (linked with many well-known benign binaries having the same false detection), while the overall detection rate was 21,7\%. Our results clearly illustrate the inefficiency of static methods in detecting the most simple malicious samples, even without any attempt to hide them. 
Figure \ref{fig:boxplots} illustrates the variation of our samples in terms of the number of sections, threads, loaded DLLs, number of functions, and size. The figure clearly showcases that while all samples perform the same tasks and are all PE executables, structurally, they are radically different. The latter is also proven by the fact that even in terms of functions, there is even greater variation. More precisely, the number of functions ranges from 6 to 81,793; note that the figure is on a logarithmic scale to illustrate the results better. 

\subsection{Open Source Static Analysis Tools}
  
In this part of the experiment, we utilize \emph{capa}~\cite{capa}, a robust, open-source capability-analysis tool developed by Mandiant widely used in cybersecurity environments, particularly within incident response teams, security operations centers (SOCs), and threat intelligence units. It can extract features from files, such as strings, disassembly, and control flow, and find combinations of features that are expressed in a common rule format. To gain ground truth, we first ran capa on the Assembly and C samples as they were the least bloated and straightforward and identified that a combination of the capa rules: \texttt{allocate or change RWX memory}, \texttt{create thread}, and \texttt{spawn thread to RWX shellcode} was able to correctly identify the shellcode execution basic block(s). Regarding the reverse Powershell payload, \texttt{execute command},\texttt{create process on Windows}, or \texttt{accept command line arguments} were the rules that indicated system command invocation. Since some of these rules can be flagged even if they are harmless, as they may be just legal procedures inside the executables, we also verify them. For each sample, we check the reported address from capa with a debugger to determine whether it actually pointed to our malicious code, eliminating false positives. For example, the Haskell binary may report just \texttt{allocate or change RWX memory}, yet this was not for our malicious code. What is interesting is how well the results from VirusTotal correlate with the results from capa. Especially in the case of shellcode samples, we have an almost one-on-one correlation with the evasive samples, indicating that some unique structural characteristics let those samples go undetected.

\begin{table*}[!th]
    \centering
    \rowcolors{1}{}{gray!10}
    \tiny
    \begin{filecontents*}{VirusTotalResultswithCapaOld.csv}
Language,Compiler,VT1,DetectionSig1,CapaDetection1,VT2,DetectionSig2,CapaDetection2
Ada,GNAT,1/70,fp,no,23/73,found,yes
Assembly,YASM/Golink,9/68,found,yes,29/68,found,yes
AutoHotKey,Ahk2EXE,9/68,found,yes,5/72,found,no
AutoIt,Au2EXE,12/70,found,yes,32/69,found,yes
C,DMC,4/69,found,yes,22/71,found,yes
C,TinyC,5/70,found,yes,45/72,found,yes
C,BCC,6/68,found,yes,21/70,found,yes
C,mingw/gcc,22/72,found,yes,51/73,found,yes
C,msvc/cl,17/73,found,yes,37/73,found,yes
C\#,bflat,7/71,found,yes,1/70,fp,no
C\#,msc,0/69,not found,yes,21/73,found,yes
C\#,csc,1/73,fp,yes,5/73,found,yes
C++,cl,17/70,fp,no,34/73,found,yes
C++,icl,17/70,fp,no,17/73,found,yes
C++,g++,5/73,found,yes,36/73,found,yes
Clojure,graal-vm,0/73,not found,no,15/73,found,no
CommonLisp,sbcl,0/72,not found,no,0/72,not found,no
Crystal,crystal,3/73,fp,no,15/73,found,yes
D,dmd,5/66,fp,no,6/73,found,no
Dart,dart,0/70,not found,no,5/69,found,no
Eiffel,ec,0/67,not found,no,11/68,found,yes
F\#,fsharpc,3/71,fp,yes,22/72,found,yes
Fortran,ifort,3/76,fp,no,17/72,found,yes
GnuCobol,cobc,4/72,found,yes,23/73,found,yes
Golang,go,4/70,found,no,16/69,found,no
Groovy,Launch4j,2/66,fp,yes,4/62,found,no
Haskell,GHC,0/71,not found,no,1/66,fp,no
IronPython,ipyc,2/67,fp,yes,2/67,fp,no
Java,graal-vm,1/73,fp,no,2/73,fp,no
Javascript,deno,0/65,not found,no,0/68,not found,no
Jscript,jsc,2/67,fp,no,16/73,found,yes
Kotlin,graal-vm,2/63,fp,no,1/73,fp,no
Kotlin,kotlin-native,0/68,not found,no,1/67,fp,no
Lua,luastatic,1/69,fp,yes,14/72,found,no
Nim,nim,0/70,not found,no,25/69,found,no
ObjectiveC,gcc,2/68,fp,yes,25/69,found,no
Pascal,fpc,0/66,not found,yes,11/66,found,yes
Perl,par,3/70,fp,no,1/71,fp,no
Phix,phix,10/72,found,no,21/67,found,no
PureBasic,pbcompiler,1/68,fp,yes,23/67,found,yes
Python,pyinstaller,6/67,found,no,3/68,fp,no
Python,nuitka,0/69,found,no,5/71,found,no
Racket,raco,0/64,not found,no,1/64,fp,no
Red,red,16/69,found,yes,22/66,found,yes
Ruby,ocra/aibica,26/68,found,no,2/71,fp,no
Rust,rustc,0/71,not found,no,16/72,found,no
Scala,graal-vm,0/73,fp,no,1/73,fp,no
Scala,launch4j,4/67,fp,no,5/63,found,no
VB .NET,vbc,5/69,found,yes,13/70,found,yes
Zig,zig,0/73,not found,no,19/68,found,yes
\end{filecontents*}
\pgfplotstabletypeset[
        col sep=comma,
        string type,
        header=true,
        every head row/.style={
            before row=\toprule,
            after row=\midrule,
            column type={>{\raggedright\arraybackslash}p{2cm}}
        },
        every last row/.style={after row=\bottomrule},
        columns={Language,Compiler,VT1,DetectionSig1,CapaDetection1,VT2,DetectionSig2,CapaDetection2},
        display columns/0/.style={column type=l, column name=\textbf{Language}},
        display columns/1/.style={column type=l, column name=\textbf{Compiler/Packager}},
        display columns/2/.style={column type=c, column name=\textbf{VT1}},
        display columns/3/.style={
            column type=c,
            column name=\textbf{Detection Sig 1},
            postproc cell content/.append style={
                @cell content=\MapFound{##1}
            }
        },
        display columns/4/.style={
            column type=c,
            column name=\textbf{Capa Detection 1},
            postproc cell content/.append style={
                @cell content=\MapCapa{##1}
            }
        },
        display columns/5/.style={column type=c, column name=\textbf{VT2}},
        display columns/6/.style={
            column type=c,
            column name=\textbf{Detection Sig 2},
            postproc cell content/.append style={
                @cell content=\MapFound{##1}
            }
        },
        display columns/7/.style={
            column type=c,
            column name=\textbf{Capa Detection 2},
            postproc cell content/.append style={
                @cell content=\MapCapa{##1}
            }
        }
    ]{VirusTotalResultswithCapaOld.csv} 
    \caption{VirusTotal and Capa results for various programming languages and compilers/packagers.}
    \label{tab:VT_results}
        \vspace{-2em}
\end{table*}

\begin{table}[!th]
    \centering
    \rowcolors{1}{}{gray!10}
    \scriptsize
    \begin{filecontents*}{shellcode_fragmentation.csv}
Language,Compiler,fragmentation,section stored,matched ratio
Ada,GNAT,none,.rdata,1
Assembly,YASM/Golink,none,.data,1
AutoHotKey,Ahk2EXE,N/A,N/A,N/A
AutoIt,Au2EXE,N/A,N/A,N/A
C,DMC,none,CRT\$XIA,1
C,TinyC,medium,.text,1
C,BCC,none,.data,1
C,mingw/gcc,none,.rdata,1
C,msvc/cl,none,.data,1
C\#,bflat,none,.rdata,1
C\#,msc,none,.sdata,1
C\#,csc,none,.text,1
C++,cl,medium,.text,1
C++,icl,none,.rdata,1
C++,g++,none,.rdata,1
Clojure,graal-vm,none,.svm\_hea,1
CommonLisp,sbcl,N/A,N/A,N/A
Crystal,crystal ,heavy,.rdata,0.86
D,dmd,heavy,.text,0.93
Dart,dart,heavy,.text,0.62
Eiffel,ec,medium,.text,1
F\#,fsharpc,heavy,.text,0.31
Fortran,ifort,none,.data,1.0
GnuCobol,cobc,none,.rdata,1.0
Golang,go,none,.rdata,1.0
Groovy,Launch4j,N/A,N/A,N/A
Haskell,GHC,N/A,N/A,N/A
IronPython,ipyc,medium,.text,1
Java,graal-vm,medium,.text,1
Javascript,deno,N/A,N/A,N/A
Jscript,jsc,medium,.text,1
Kotlin,graal-vm,medum,.text,1
Kotlin,kotlin-native,medium,.text,1
Lua,luastatic,N/A,N/A,N/A
Nim,nim,none,.data,1
ObjectiveC,gcc,none,.text,1
Pascal,fpc,medium,.text,1
Perl,par,N/A,N/A,N/A
Phix,phix,medium,.text,1
PureBasic,pbcompiler,none,.data,1
Python,pyinstaller,N/A,N/A,N/A
Python,nuitka,N/A,N/A,N/A
Racket,raco,N/A,N/A,N/A
Red,red,none,.data,1
Ruby,ocra/aibica,N/A,N/A,N/A
Rust,rustc,heavy,.rdata/.text,1
Scala,graal-vm,medium,.text,1
Scala,launch4j,N/A,N/A,N/A
VB .NET,vbc,medium,.text,1
Zig,zig,none,.text,1

\end{filecontents*}
\pgfplotstabletypeset[
        col sep=comma,
        string type,
        header=true,
        every head row/.style={
            before row=\toprule,
            after row=\midrule,
            column type={>{\raggedright\arraybackslash}p{2cm}}
        },
        every last row/.style={after row=\bottomrule},
        columns={Language,Compiler,fragmentation,section stored,matched ratio},
        display columns/0/.style={column type=l, column name=\textbf{Language}},
        display columns/1/.style={column type=l, column name=\textbf{Compiler/Packager}},
        display columns/2/.style={column type=c, column name=\textbf{Fragmentation}},
        display columns/3/.style={column type=c, column name=\textbf{Section Stored}},
        display columns/4/.style={column type=c, column name=\textbf{Matched Ratio}},
    ]{shellcode_fragmentation.csv} 
    \caption{Shellcode fragmentation through pattern matching on binaries.}
    \label{tab:shellcode_fragmentation}
        \vspace{-2em}
\end{table}

\subsection{Shellcode Fragmentation}

To assess how immune the binaries produced by different programming languages are to shellcode pattern matching, and since there is no intended obfuscation of the payloads, we conducted an experiment utilizing a custom-developed pattern-matching script to analyze the bytes from the raw binaries on disk. We allowed the matching operation to search for chunks of shellcode by fine-tuning two parameters for each binary, namely \emph{Maximum Gap} and \emph{Minimum Chunk Size}. For the former, we set the maximum allowed gap between matched shellcode bytes to 60, allowing flexibility for scattered patterns. For the latter, only matched sequences of at least 4 bytes were considered valid, reducing false positives from incidental matches of very small byte sequences and returning the sequences with the highest matching ratio.

We also performed pattern matching in the reversed order of bytes to identify possible stack-based shellcodes (for example, 
\lstinline[language={[x86masm]Assembler},columns=fixed]{push <byte>}
). Since the objective of this experiment was to statically locate the raw dummy payload using static methods and not actually locate the shellcode by any means necessary, we did not use any dynamic analysis tools.

After executing the script, all identified patterns were manually reviewed using a debugger and a hex editor to confirm the matches and eliminate false positives. This step ensured that only genuine shellcode patterns were considered in the results. Table \ref{tab:shellcode_fragmentation} categorizes the matches into four levels of fragmentation, namely:
\begin{enumerate*}
    \item \textbf{None}: Shellcode bytes were sequential, indicating that there was no fragmentation;
    \item \textbf{Medium}: Shellcode bytes were scattered but with gaps within a range; 
    \item \textbf{Heavy}: Shellcode bytes were fragmented with scattered chunks of large distance, wherein each chunk bytes was sequential or had fixed gaps within a range;
    \item \textbf{N/A}: The script was unable to confidently identify the shellcode in the binary, indicating the highest level of fragmentation or potential complex encoding.
\end{enumerate*}

The results showed considerable discrepancies in pattern matching; for example, samples written in languages such as C and C++ retained, usually all shellcode bytes in sequential order or had a fixed gap between the bytes, leading to relatively straightforward detection. However, other languages demonstrated significant byte fragmentation and variations in memory layout, complicating static detection. For instance, our Rust implementation showed a complex pattern with the shellcode bytes dispersed irregularly throughout the binary at various offsets (e.g., starting with an initial block of 192 bytes at \texttt{0x16570} with no gaps, followed by a smaller, non-sequential block with gaps of up to 13 bytes at \texttt{0x4ee} and continued again with non-continuous blocks of shellcode at address \texttt{0x16630}). In the unique case of Phix, the shellcode was pushed on the stack byte by byte. Finally, in languages like Lisp and Haskell, we could not find any part of the shellcode with confidence. Another interesting result is that most of the samples with low detection rates also had their shellcode pattern unidentified within the binary, indicating another correlation between their structure and evasive behavior.

 \begin{table*}[!th]
    \centering
    \rowcolors{1}{}{gray!10}
    \tiny
    \begin{filecontents*}{reverse_engineering_metrics.csv}
Language,Compiler,functions,funcs exec,avg func size,bb hits,inst hits,CC,ind jmps,ind calls,threads
Ada,GNAT,1695,92,171.08,493,2482,3.51,8,36,3
Assembly,YASM/Golink,5,5,19,5,26,1,0,0,2
AutoHotKey,Ahk2EXE,1464,147,1169.82,3606,15128,48.44,23,12,4
AutoIt,Au2EXE,2282,132,287.77,378,8441,9.65,0,44,6
C,DMC,69,34,106.53,186,902,4.94,0,0,3
C,TinyC,15,10,215.3,10,500,1,0,0,2
C,BCC,309,65,101.14,69,783,3.16,0,1,3
C,mingw/gcc,79,13,98.24,18,482,4.03,0,5,2
C,msvc/cl,436,47,129.43,91,1061,4.66,2,0,2
C\#,bflat,3718,349,166.68,683,9769,4.43,6,16,4
C\#,csc,17736,784,142.76,440,4354,8052,36,0,8
C++,cl,343,26,141.3,6,392,3.81,0,0,3
C++,icl,451,37,161.54,74,993,5.17,0,0,3
C++,g++,79,33,98.24,93,445,4.03,0,5,3
Clojure,graal-vm,7314,1042,1284.87,13436,133483,31.32,7,564,4
CommonLisp,sbcl,781,195,560,2087,26931,134.4,1,101,5
Crystal,crystal,3327,193,203.16,586,5682,6.98,4,6,4
D,dmd,2409,1429,164.5,720,10982,4.13,5,32,2
Dart,dart,9251,916,308.88,2167,40830,6.86,13,141,1
Eiffel,ec,4051,762,146.58,894,18068,2.97,0,4,2
Fortran,ifort,914,291,492.85,2183,11009,17.75,21,1,3
GnuCobol,cobc,100,22,95.8,45,227,2.90,0,0,1
Golang,go,1616,439,382.97,4478,35007,1.77,2,21,6
Groovy,Launch4j,162,130,131.31,364,4068,4.92,0,1,2
Haskell,GHC,2974,2318,187.3,2200,22596,4.97,276,47,6
Java,graal-vm,6969,996,969.05,12764,125244,23.35,6,413,4
Javascript,deno,81792,1717,460.99,37475,280860,9.75,1521,0,5
Kotlin,graal-vm,6902,981,973.44,12955,55424,23.4,5,431,4
Kotlin,kotlin-native,1574,206,150.85,574,10582,4.67,3,26,2
Lua,luastatic,1545,332,350.55,2821,16246,10.16,54,29,3
Nim,nim,359,130,226.28,309,5343,2.26,0,24,2
ObjectiveC,gcc,52,24,113.2,43,291,2.27,0,0,2
Pascal,fpc,429,145,128.86,305,4051,3.77,0,41,3
Perl,par,2821,82,146.79,276,15570,4.71,5,431,3
Phix,phix,167,82,522.39,390,1842,22.46,0,11,3
PureBasic,pbcompiler,44,10,36.30,2,113,1.10,1,0,2
Python,pyinstaller,819,117,302.7,577,6075,10.77,4,22,2
Python,nuitka,370,79,670.63,1234,5841,12.92,3,19,2
Racket,raco,116,49,148.71,328,2219,4.51,0,49,2
Red,red,22,8,99.0,13,224,1.25,0,0,3
Ruby,ocra/aibica,132,63,234.63,488,3077,5.98,0,48,2
Rust,rustc,337,36,103.5,95,595,2.42,2,4,2
Scala,graal-vm,7021,967,1019.57,13186,130330,23.61,5,433,4
Scala,launch4j,167,116,142.51,432,4050,4.79,0,1,2
Zig,zig,639,212,374.8,1191,10269,2.05,4,11,2
\end{filecontents*}
\pgfplotstabletypeset[
        col sep=comma,
        string type,
        header=true,
        every head row/.style={
            before row=\toprule,
            after row=\midrule,
            column type={>{\raggedright\arraybackslash}p{2cm}}
        },
        every last row/.style={after row=\bottomrule},
        columns={Language,Compiler,functions,funcs exec,avg func size,bb hits,inst hits,CC,ind jmps,ind calls},
        display columns/0/.style={column type=l, column name=\textbf{Language}},
        display columns/1/.style={column type=l, column name=\textbf{Compiler}},
        display columns/2/.style={column type=c, column name=\textbf{\#Func}},
        display columns/3/.style={column type=c, column name=\textbf{\#Func Exec}},
        display columns/4/.style={column type=c, column name=\textbf{Avg Func Size}},
        display columns/5/.style={column type=c, column name=\textbf{\#BB Hits}},
        display columns/6/.style={column type=c, column name=\textbf{\#Instr Hits}},
        display columns/7/.style={column type=c, column name=\textbf{CC}},
        display columns/8/.style={column type=c, column name=\textbf{\#Ind Jmps}},
        display columns/9/.style={column type=c, column name=\textbf{\#Ind Calls}},
    ]{reverse_engineering_metrics.csv} 
    \caption{Reverse engineering metrics I.}
    \label{tab:reverse_engineering_metrics}
        \vspace{-2em}
\end{table*}

 \begin{table}[!th]
    \centering
    \rowcolors{1}{}{gray!10}
    \tiny
    \begin{filecontents*}{main_module_statistics.csv}
language,num_nodes,num_edges,total_traversals,indirect_calls,indirect_jumps,normalized_entropy
Ada,44,45,74,63,12,0.98
Assembly,0,0,0,0,0,0
AutohotKey,35,64,4973,1403,3571,0.66
AutoIt,44,73,5983,1678,4305,0.57
C-bcc,1,1,21,0,52,0
C-cl,2,2,3,0,4,0.91
C-gcc,5,4,4,5,0,1.0
C-tcc,0,0,0,0,0,0
C-dmc,0,0,0,0,0,0
C\#-bflat,22,34,24559,329,24321,0.53
C\#-csc,36,127,237,0,237,0.43
C++-cl,0,0,0,0,0,0
C++-icl,0,0,0,0,0,0
C++-g++,5,4,4,5,0,1.0
Clojure,571,890,19176,18853,324,0.53
CommonLisp,102,126,706,693,14,0.54
Crystal,10,18,2088,1032,1057,0.30
D,37,53,199,186,14,0.58
Dart,154,249,34673,14750,19924,0.41
Eiffel,4,7,41,42,0,0.74
Fortran,22,26,55,1,55,0.93
GnuCobol,0,0,0,0,0,0
Golang,23,69,6219,6057,163,0.34
Groovy,1,0,0,1,0,0
Haskell,323,652,8265,488,7778,0.66
Java,419,634,19879,19732,263,0.57
Javascript,1521,3427,403815,403815,0,0.56
Kotlin-graalvm,436,660,19917,19657,261,0.56
Kotlin-native,29,41,189,187,3,0.63
Lua,83,220,3753,869,2885,0.57
Nim,24,30,35,36,0,0.98
ObjC,0,0,0,0,0,0
Pascal,41,56,88,89,0,0.96
Perl,436,660,19917,19657,261,0.56
Phix,11,25,30967,30968,11,0.24
PureBasic,1,0,0,0,1,0
Python-pyinstaller,26,37,563,453,110,0.51
Python-nuitka,22,27,76,38,39,0.89
Racket,49,70,795,796,0,0.62
Red,0,0,0,0,0,0
Ruby,48,89,432,432,0,0.65
Rust,6,6,6,5,2,1.0
Scala-graalvm,438,669,20207,19945,263,0.55
Scala-launch4j,1,0,0,1,0,0
Zig,15,24,171,17,155,0.50
\end{filecontents*}
\pgfplotstabletypeset[
        col sep=comma,
        string type,
        header=true,
        every head row/.style={
            before row=\toprule,
            after row=\midrule,
            column type={>{\raggedright\arraybackslash}p{2cm}}
        },
        every last row/.style={after row=\bottomrule},
        columns={language,num_nodes,num_edges,total_traversals,indirect_calls,indirect_jumps,normalized_entropy},
        display columns/0/.style={column type=l, column name=\textbf{Language}},
        display columns/1/.style={column type=l, column name=\textbf{\#Nodes}},
        display columns/2/.style={column type=l, column name=\textbf{\#Edges}},
        display columns/3/.style={column type=l, column name=\textbf{\#Traversals}},
        display columns/4/.style={column type=l, column name=\textbf{\#Tot. Ind Cals}},
        display columns/5/.style={column type=l, column name=\textbf{\#Tot. Ind Jmps}},
        display columns/6/.style={column type=l, column name=\textbf{CFG Entropy}},   
    ]{main_module_statistics.csv} 
    \caption{Reverse engineering metrics II.}
    \label{tab:reverse_engineering_metrics2}
        \vspace{-2em}
\end{table}

\subsection{Reverse Engineering Metrics}

In this section, we try to measure the shellcode binaries from the runtime complexity perspective. Although reverse engineering difficulty is not easy to measure as it is heavily based on the human element of intuition and expertise as well as on how fine-tuned the used tools are, our high-level metrics indicate a connection between our most evasive samples and their actual complexity. In particular, we focused on the following key metrics (see Tables~\ref{tab:reverse_engineering_metrics} and \ref{tab:reverse_engineering_metrics2}): number of functions, number of functions actually executed, average function size of executed functions, unique basic blocks executed, unique instructions executed based on the address they were found, meaning that if a particular instruction of a basic block is traversed more than once, it is not counted. 

We also calculated the average cyclomatic complexity of the executed functions, the unique indirection calls and jumps executed, as well as the number of threads spawned. To acquire our results, we collected complete instruction traces of the executables with the help of IDAPro and its PinTracer debugger without taking into account traces from Windows dynamic libraries. We intentionally focused on indirect jumps and calls while excluding other control flow operands (for example, returns). This decision was motivated by the desire to capture the control flow aspects that most significantly impact program complexity and dynamic behavior and create static analysis challenges. Indirect control flow transfers, such as indirect jumps and indirect calls, are crucial in representing dynamic behavior in programs. They occur when the target of a jump or call is determined at runtime, often through function pointers, virtual method tables, or dynamic dispatch mechanisms. 

By concentrating on indirect calls and jumps, we essentially measure how much of the program's control flow is determined at runtime rather than at static analysis time. A high number or a dense network of indirect branches can suggest more dynamic behavior, code unintended obfuscation techniques, or pointer-based dispatch tables, all of which add to reverse engineering difficulty. To that end, we also constructed CFGs that capture the indirection aspect of the shellcode samples where each node represented an indirect \texttt{jmp} or \texttt{call} and measure the total traversals, which indicate how many indirections occurred in total during execution. A large number of edge traversals can imply that the program frequently relies on indirect control flow to reach various parts of the code, suggesting that any attempt at reverse engineering must continuously resolve these runtime-dependent branches. We further incorporate an information-theoretic measure; Shannon's entropy to capture the unpredictability of edge transitions. 

Entropy, computed from edge frequencies, quantifies how the transitions are distributed among possible indirect edges. High entropy indicates that the program does not favor a small set of indirect branches but rather exercises many of them with similar frequency, increasing uncertainty for the analyst. On the contrary, a low-entropy graph may be complex in structure but predictable in practice if only a few edges are predominantly taken.  

We also need to mention that the sizes of the trace log files ranged from a few kilobytes to almost 10 Gigabytes. We also did not include in the analysis the .NET languages except for the C\#-csc sample, which was the only sample that included the runtime environment, since usually, .NET compiled languages do not need to include the runtime environment to be able to run in a target. Also, these samples can be trivially reversed using dnSpy and .NET-focused tools.

We observed that in almost all cases that reported low detection scores, there were many indirections or large and complex functions, showcasing how the runtime environment of each language adds vast amounts of complexity to simple malicious code.

\subsection{Case Study: Haskell Reverse Engineering}

In this section, we chose to investigate the challenges posed by reverse engineering one of our shellcode samples that had evasive behavior, in particular the Haskell executable, focussing on how the inherent characteristics of the GHC runtime and its execution model complicate traditional analysis techniques compared to the corresponding C sample compiled with the Windows MSVC toolchain. Our goal is to highlight some of the key differences we observed that introduce substantial complexity and how static disassembly or debugging struggles to provide accurate insights into their execution. However, providing a complete analysis of the binary is beyond the scope of this work, as this would require a deep dive into the GHC runtime and Lambda calculus.

In C, control flow is generally direct and imperative. Each instruction or function call follows in a predictable sequence, and system calls, such as memory allocation through \texttt{VirtualAlloc}, are explicit and immediately visible. For instance, in our sample, the first thing an analyst would see in the disassembled code is that the steps are linear (see Listing~\ref{disasC}).

\begin{lstlisting}[language={[x86masm]Assembler},label={disasC},caption={Disassembled snippet of C shellcode sample.}]
lea     rax, [rsp+188h+var_148]
lea     rcx, unk_7FF7EDACB000
mov     rdi, rax
mov     rsi, rcx
mov     ecx, 115h
rep movsb
mov     r9d, 40h ; '@'  ; flProtect
mov     r8d, 3000h      ; flAllocationType
mov     edx, 115h       ; dwSize
xor     ecx, ecx        ; IP address
call    cs:VirtualAlloc
mov     [rsp+188h+lpStartAddress], rax
mov     r8d, 115h
lea     rdx, [rsp+188h+var_148]
mov     rcx, [rsp+188h+lpStartAddress]
call    sub_7FF7EDABFBA0
\end{lstlisting}

Furthermore, since the disassembled code is straightforward, it can be analyzed statically in a few minutes. In (Listing~\ref{disasC}), the shellcode is loaded from the section in which it resides (.data) in 
\texttt{rcx}, and then it gets copied in the stack space byte by byte. Then, \texttt{VirtualAlloc} allocates some space, and the actual shellcode gets copied from the stack to the fresh \texttt{RWX} allocated space. 

On the other hand, reversing Haskell binaries presents significant challenges due to the intricacies of its execution model. Going from the call to \texttt{VirtualAlloc} to copying shellcode into the allocated space involves more than 100 thousand instructions based on the execution trace we got as opposed to the C sample, which was only five instructions. Throughout the disassembled code, user code is blended with the STG-machine code that handles each own stack and heap, has sophisticated pointer management and garbage collector, and also makes heavy use of indirection jumps because of its lazy evaluation that defers computation until needed and continuation-passing. 

A continuation is a callback function that expects the result of a previous computation as an argument; in other words, it represents `what to do next.' Closures and continuations are among the main reasons Haskell reported such a high number of indirections.
Here, the shellcode is initialized from the raw binary written in heap memory, then prepared for the interaction with the FFI, and inherits a unique obfuscation scheme to execute the shellcode. The code in Listing~\ref{disasHaskell} shows how each byte of the shellcode is stored in the raw executable and what is executed during the initialization of our malicious shellcode in heap memory. Considering that \texttt{rbp} and \texttt{r12} are the equivalent of stack and heap registers in STG-machine, the code after a series of memory checks goes through a memory allocation routine using \texttt{newCAF} and \texttt{allocateMightFail} GHC functions. Finally, the instruction \texttt{mov qword ptr [r12], 0FCh} stores the first byte of our shellcode (\texttt{0xFC}) at the address pointed to by \texttt{r12} in heap memory. During what we just described, many other procedures occur, such as thread handling and garbage collection, making the code even more incomprehensible than the Assembly produced by C.

As we saw, the executable inherits from the actual language runtime an obfuscation scheme where the shellcode is stored and loaded dynamically byte by byte and is only fully assembled in executable memory at runtime.

\begin{lstlisting}[language={[x86masm]Assembler},label={disasHaskell},caption={Disassembled snippet of Haskell shellcode sample.}]
lea     rax, [rbp-20h]
cmp     rax, r15
jb      short loc_40BE86
add     r12, 10h
cmp     r12, [r13+358h]
ja      short loc_40BE7B
sub     rsp, 8
mov     rcx, r13
mov     rdx, rbx
sub     rsp, 20h
xor     eax, eax
call    newCAF
add     rsp, 28h
test    rax, rax
jz      short loc_40BE79
mov     qword ptr [rbp-10h], 43BFF8h
mov     [rbp-8], rax
mov     qword ptr [r12-8], 4C01C8h
mov     qword ptr [r12], 0FCh
lea     rax, [r12-7]
mov     r14d, offset base_GHCziWord_zdfNumWord8_closure
mov     qword ptr [rbp-20h], 43D1A0h
mov     [rbp-18h], rax
add     rbp, 0FFFFFFFFFFFFFFE0h
jmp     base_GHCziNum_fromInteger_info
\end{lstlisting}

\section{Discussion}

Languages such as Java, Clojure, Scala, Kotlin, and JavaScript, which embed substantial runtimes or rely on JIT compilation, consistently produced large, complex binaries. These executables exhibited extensive CFGs (high node/edge counts), numerous indirect calls/jumps, and large numbers of functions.
VirusTotal results showed that such complexity often correlated with higher detection rates or initial false positives. Heuristic-based detection engines frequently flagged these binaries as suspicious, likely due to unfamiliar or intricate patterns in control flow and the presence of runtime scaffolding code. Although subsequent passes or capa reports sometimes clarified these detections, the initial suspicion underscores the challenges static AV tools face when analyzing runtime-heavy executables.

In contrast, binaries produced by traditional compiled languages (C, Fortran, Ada) and straightforward compilers tended to have simpler structures. With fewer functions, less fragmentation, and minimal indirect control flows, these binaries were more transparently analyzable. Their matched ratios were often perfect (1.0), indicating easy alignment between the binary and static analysis tools.
As a result, detection outcomes were more predictable. Such binaries were either not detected at all or consistently identified as benign. When detections occurred, they were more easily interpreted, reducing the likelihood of persistent false positives.

Heavy fragmentation corresponded to lower matched ratios, complicating static analysis and potentially increasing false-positive rates. Fragmented code segments impeded effective disassembly and structured understanding of the binary. As a result, AV engines that rely on pattern matching or heuristic scanning may misinterpret such binaries as suspicious, even without known malicious signatures.

The prominence of indirect calls and jumps in runtime-heavy languages serves as an additional complexity signal. Indirect branching complicates the control-flow analysis, challenging both AV signatures and CFG extraction tools. The correlation between indirect control-flow patterns and AV detections or FPs suggests that complexity in flow redirection can raise the heuristic suspicion threshold and lead to detections.

Finally, normalized entropy provided insights into the uniformity of byte distributions. High entropy often occurs in packed or obfuscated binaries, which can appear anomalous to AV engines as most modern malware uses some packer. While not the sole predictor of detection outcomes, elevated entropy combined with fragmentation and complex control-flow patterns often coincided with uncertain or cautious AV responses.

Our results highlight that no single metric conclusively determines AV detection outcomes. Instead, a combination of factors—runtime complexity, fragmentation, control-flow intricacy, entropy, and function-level distributions—influences how AV engines classify binaries. Therefore, from the defender's and analysts' perspectives, understanding these correlations, creating more robust signatures, and extending the scope of tools to consider more programming languages and compilers is imperative, as threat actors can easily exploit this gap. From an attacker's perspective, the findings indicate that complexity and indirect control flow can serve as evasive techniques, potentially raising false alarms or complicating detection. However, sustained complexity may also attract scrutiny, highlighting a delicate balance between obfuscation and detection risk.

\section{Conclusion}

Malware is predominantly written in C/C++ and is compiled with Microsoft's compiler. However, trying to answer \textbf{RQ1} with our experiments, our work practically shows that by shifting the codebase to another, less used programming language or compiler, malware authors can \emph{significantly decrease the detection rate} of their binaries but simultaneously \emph{increase the reverse engineering effort} of the malware analysts. It is crucial to note that the malware authors do not necessarily need to radically change their codebase, as, for instance, the choice of another compiler, even for famous programming languages like C, can have the same impact. Our experimental results illustrate that there are significant deviations in how programming languages and compilers generate binaries, and that they can serve as an additional layer of obfuscation for malware authors.

The root cause for the disparities that we raise (\textbf{RQ2}), as highlighted with our use case in Haskell and the metrics for each tested pair of programming language and compiler, is that there are radically different ways that each of them reaches the same result. For instance, different ways of storing strings and different approaches in the internal representation of functions can render many static detection rules useless. As a result, there is no "one-size-fits-all" approach, so further research is necessary to systematically identify these differences and group them. 

Moreover, answering \textbf{RQ3}, this shift may come with additional benefits for attackers. An obvious case is cross-compilation and multi-platform targeting languages, which enable malware authors to build a single malware variant and have it compiled for multiple operating systems. This strategy can significantly reduce the time and number of tools needed to achieve their objectives, thereby expanding the scope of any hostile campaign. IoT devices, in particular, support a range of CPU environments, making it necessary for malware targeting these devices to be compatible with not only x86 and x64 architectures but also various other architectures such as ARM, MIPS, m68k, SPARC, and SH4. 

A typical example is Mirai ~\cite{antonakakis2017understanding}, which uses GCC, yet one of its successors, NoaBot~\cite{noabot}, uses uClibc-based cross-compiler and is statically built to target embedded Linux systems. 
In this regard, other options could be more efficient. For instance, Go can be cross-compiled to all major operating systems, as well as Android, JavaScript, and WebAssembly. One of its advantages is that it provides statically compiled binaries by default, eliminating runtime dependencies and simplifying deployment on target systems. Go also features a robust package ecosystem that allows developers to easily pull in code from other sources. In general, cross-compilation in Go is as simple as setting two environment variables, making it almost trivial to modify the build process to produce binaries for every major platform. As a result, malware can be developed at a faster rate, targeting a broader range of architectures and systems. Indeed, HinataBot~\cite{hinata}, a descendant of Mirai, is developed in Go to take advantage of the above. The HinataBot was more difficult to be discovered by detection systems. Unfortunately, the bar to creating a new variant of Mirai using Go or other languages is low, and criminal groups make their own variations~\cite{Mirai-evolution}.

Beyond cross-compilation, there are several other reasons to witness more changes in the malware codebase. After all, malware developers, like any other software engineers, have specific needs when choosing programming languages and tools. Different languages offer various benefits for different scenarios, and the choice of language can significantly impact the development and functionality of malware. For instance, built-in security mechanisms and type safety may be prioritized by ransomware authors who want to avoid leaks of the encryption keys to guarantee that their victims will not be able to develop decryptors. A typical example is Rust, which offers built-in memory mechanisms to prevent common vulnerabilities and type safety. Other aspects can include library availability; facilitating interaction with the underlying operating system and enabling critical malware functions, low-level access, and control over memory layout; having full control over the malware's behavior and performance but also direct compilation to machine code; creating an executable file directly and use other tools for obfuscation.

While shifting to another programming language may seem complicated, especially when considering less popular ones, large language models (LLMs) may come to the rescue; after all, they have proven their capacity in generating code quite accurately~\cite{ouyang2023llm,liu2024your,10507163,10698405,guo2024esorics} and various cybersecurity tasks~\cite{deng2023pentestgpt,PatsakisCL24}, and malicious actors are abusing them. As a result, they can translate code from one programming language to another, requiring little fine-tuning. This way, malware authors can seamlessly develop loaders, droppers, and other components in languages they may not be familiar with.

It is true that the malware that we examine in this work represents a small fragment of the total; nevertheless, it is stealthier and introduces more bottlenecks for the reverse engineer. Given that the APT groups are shifting their codebases and the malware-as-a-service model facilitates the trading of malware so different malware mixtures per campaign can be purchased, this diversification is expected to continue. By disregarding these samples and only focusing on traditional programming languages and compilers, we provide malware authors with an effective hideout that they can easily exploit. Therefore, we believe that a deeper analysis of the executables produced by other compilers and programming languages is needed to improve detection rates but also develop better reverse engineering tools.

\section*{Acknowledgment}
This work was supported by the European Commission under the Horizon Europe Programme as part of the project SafeHorizon (Grant Agreement no. 101168562). The content of this article does not reflect the official opinion of the European Union. The responsibility for the information and views expressed therein lies entirely with the authors.

\end{document}